\documentclass[twocolumn,times]{aastex62}
\received{XXXX XX, 201X}
\revised{XXXX XX, 201X}
\accepted{\today}
\submitjournal{ApJ}

\shorttitle{Delay}
\shortauthors{Czerny et al.}

\begin{document}

\title{\large Time delay measurement of Mg II line in CTS C30.10 with SALT}

\correspondingauthor{B. Czerny}
\email{bcz@cft.edu.pl}

\author{Bo{\.z}ena Czerny}
\affiliation{Center for Theoretical Physics, Polish Academy of Sciences, Al. Lotnikow 32/46, 02-668 Warsaw, Poland}

\author{Aleksandra Olejak}
\affiliation{Center for Theoretical Physics, Polish Academy of Sciences, Al. Lotnikow 32/46, 02-668 Warsaw, Poland}

\author{Mateusz Ra\l owski}
\affiliation{Center for Theoretical Physics, Polish Academy of Sciences, Al. Lotnikow 32/46, 02-668 Warsaw, Poland}

\author{Szymon Koz\l owski}
\affiliation{Warsaw University Observatory, Al. Ujazdowskie 4, 88-478 Warsaw, Poland}

\author{Mary Loli Martinez Aldama}
\affiliation{Center for Theoretical Physics, Polish Academy of Sciences, Al. Lotnikow 32/46, 02-668 Warsaw, Poland}

\author{Michal Zajacek}
\affiliation{Center for Theoretical Physics, Polish Academy of Sciences, Al. Lotnikow 32/46, 02-668 Warsaw, Poland}

\author{Wojtek Pych}
\affiliation{Nicolaus Copernicus Astronomical Center, Polish Academy of Sciences, ul. Bartycka 18, 00-716 Warsaw, Poland}

\author{Krzysztof Hryniewicz}
\affiliation{Nicolaus Copernicus Astronomical Center, Polish Academy of Sciences, ul. Bartycka 18, 00-716 Warsaw, Poland}

\author{Grzegorz Pietrzy\' nski}
\affiliation{Nicolaus Copernicus Astronomical Center, Polish Academy of Sciences, ul. Bartycka 18, 00-716 Warsaw, Poland}

\author{C. Sobrino Figaredo}
\affiliation{Astronomisches Institut - Ruhr Universität Bochum, Germany}

\author{Martin Haas}
\affiliation{Astronomisches Institut - Ruhr Universität Bochum, Germany}

\author{Justyna \' Sredzi\' nska}
\affiliation{Nicolaus Copernicus Astronomical Center, Polish Academy of Sciences, ul. Bartycka 18, 00-716 Warsaw, Poland}

\author{Magdalena Krupa}
\affiliation{Astronomical Observatory of the Jagiellonian University,
  Orla 171, 30-244 Cracow, Poland}

\author{Agnieszka Kurcz}
\affiliation{Astronomical Observatory of the Jagiellonian University,
  Orla 171, 30-244 Cracow, Poland}

\author{Andrzej Udalski}
\affiliation{Warsaw University Observatory, Al. Ujazdowskie 4, 88-478 Warsaw, Poland}

\author{Marek Gorski}
\affiliation{Departamento de Astronomiıa, Universidad de Concepcion, Casilla 160-C, Chile}

\author{Vladimir Karas}
\affiliation{Astronomical Institute, Academy of Sciences, Bocni II, CZ-141 31 Prague, Czech Republic}

\author{Swayamtrupta Panda}
\affiliation{Center for Theoretical Physics, Polish Academy of Sciences, Al. Lotnikow 32/46, 02-668 Warsaw, Poland}
\affiliation{Nicolaus Copernicus Astronomical Center, Polish Academy of Sciences, ul. Bartycka 18, 00-716 Warsaw, Poland}

\author{Marzena Sniegowska}
\affiliation{Center for Theoretical Physics, Polish Academy of Sciences, Al. Lotnikow 32/46, 02-668 Warsaw, Poland}

\author{Mohammad-Hassan Naddaf}
\affiliation{Center for Theoretical Physics, Polish Academy of Sciences, Al. Lotnikow 32/46, 02-668 Warsaw, Poland}

\author{Marek Sarna}
\affiliation{Nicolaus Copernicus Astronomical Center, Polish Academy of Sciences, ul. Bartycka 18, 00-716 Warsaw, Poland}



\begin{abstract}
  We report 6 yr monitoring of a distant bright quasar CTS C30.10 ($z = 0.90052$) with the Southern African Large Telescope (SALT). We measured the rest-frame time-lag of $562\pm2$ days between the continuum variations and the response of the Mg II emission line, using the Javelin approach. More conservative approach, based on five different methods, imply the time delay of $564^{+109}_{-71}$ days. This time delay, combined with other available measurements of Mg II line delay, mostly for lower redshift sources, shows that the Mg II line reverberation implies a radius-luminosity relation very similar to the one based on a more frequently studied H$\beta$ line.
\end{abstract}




\section{Introduction}
\label{sec:intro}

Broad emission lines coming from the Broad Line Region (BLR) are the most characteristic properties of most Active Galactic Nuclei (AGN). The innermost region of the accretion flow in bright AGN, including the BLR, is spatially unresolved, with the exception of the first image of the BLR in 3C 273 achieved with GRAVITY \citep{gravity2018}. Nevertheless, observations of BLR, in particular time variability, allow us to gain an insight into the structure of the nuclear region, provides a tool to measure the black hole mass, and contains a promise of possible cosmological applications.

The first image of BLR seen with GRAVITY in general supports the view which was formulated before:  BLR forms a flattened structure with the symmetry axis aligned with the jet axis, the velocity of the clouds are predominantly Keplerian, roughly consistent (within a factor 2) with the result obtained from traditional reverberation mapping by \citet{zhang2018}.

Thus reverberation mapping is a reliable method of measuring the size of the BLR and the black hole mass \citep{kaspi2000,peterson2004,mejia2018}. The mass determination requires the measurement of the time delay between a given line and a continuum, the measurement of the line width (either FWHM or $\sigma$), and the knowledge of the virial factor. Observations suggested a strong coupling between the BLR size obtained from such delay measurements and the monochromatic luminosity \citep{kaspi2000,bentz2013}, and such a generalized law leads to single-spectrum methods of black hole mass calculations in large quasar catalogs \cite[e.g.][]{shen2011}.

However, most of the measurements were done only for relatively nearby AGN, in H$\beta$ line. In the case of more distant quasars, their H$\beta$ line moves to the IR domain and optical observations provide only the UV part of the spectrum with Mg II and CIV lines. There were attempts to bridge the use
of H$\beta$ and other lines by taking into account the systematic differences in the line widths as well as due to the fact of using also a different part of the continuum as a reference \citep[see e.g.][]{woo2018}. Statistical scaling have, however, some limitations, particularly that higher redshift sources have frequently higher masses and/or Eddington ratios, so the direct confirmation of the scaling laws from reverberation mapping in lines other than H$\beta$ is important.

In this paper we present our reverberation campaign in Mg II line of a quasar CTS C30.10 ($z = 0.90052$). Mapping in Mg II is more difficult since the variability of Mg II line is in general lower than in H$\beta$ \citep{goad1999,zhu2017}. There are only a few successful measurements in this line so far. \citet{clavel1991} pioneered the reverberation mapping in the UV lines, but the constraint they received for NGC 5548 were very approximate, $\sim 34 - 72$ days for Mg II line. The first firm delays, in two epochs were measured for NGC4151 by \citet{metzroth2006}.
\citet{shen2016} reported the time delay measurement in Mg II for six objects at moderate redshift (between 0.4253 and 0.7510), and all their sources were of relatively low luminosity ($\log L_{5100}$ below 44.416). \citet{lira2018} reports the delay for the quasar CT252 ($z =  1.818$, $ \log L_{5100} = 46.48$).
\citet{cackett2015} performed monitoring of NGC 5548 in UV but they were not able to determine the Mg II time delay from their data. Our campaign lasted 6 years and we are able to report the definite delay measurement for more distant and very bright quasar ($\log L_{3000} = 46.023$). Very preliminary results of this effort were published in a short conference report \citep{czerny_pas2018}.

\section{Observations and data reduction}
\label{sect:observations}

Quasar CTS C30.10  has been found in  the  Calan-Tololo
Survey, as a part of survey aimed to find new bright quasars in the southern part of the sky. The source is located at
RA = 04h47m19.9s, Dec=-45d37m38s (J2000.0). Quasar identification has been confirmed by \citet{maza1993}. The source is bright
($V = 17.2$, as given by NED), the redshift ($z = 0.910$ in NED) has been revised to $z = 0.9000$ by \citep{modzelewska2014}.
The source has been observed with the SALT from 6 December 2012 till 10 December 2018.

\subsection{Spectroscopy}

The source was observed in a slit spectroscopy mode, using the Robert
Stobie  Spectrograph  (RSS;  \citealt{burgh2003,kobulnicky2003,smith2006}) in the
service mode. We collected 26 observations, each consisting of two observing blocks, with exposures lasting $\sim 800$ s. Slit width of 2'' was used, and
RSS PG1300 grating, which corresponds to the spectral resolution of 1047 at 5500 \AA.  The dates of the observations
are reported in Table~\ref{tab:SALT_data}.

\begin{table*}
  \caption{SALT spectroscopy}
  \label{tab:SALT_data}
  \centering                          
  \begin{tabular}{l r r r  r  r  r  r  r }        
    \hline\hline
    obs. & JD  & EW(Mg II) & err+ & err- & EW(Fe II) & err+  & err-\\
    no.& - 2 540 000 &  [\AA] & [\AA] & [\AA] & [\AA] & [\AA] & [\AA] \\
    \hline
    1 & 6268.5563 &  25.55  &    0.49  &    0.47 &  7.47 &  1.00 &  0.99\\
    2 & 6314.4354 &  26.51  &    0.66  &    0.62 & 7.59 &   1.13 &  1.16\\
    3 & 6371.2675 &  27.35  &    0.83  &    0.77 &  9.47 &  1.39 &  1.39\\
    4 & 6509.6468 &  29.68  &    0.57  &    0.54 & 11.57 &  0.95 &  0.95\\
    5 & 6722.3211 &  29.65  &    0.66  &    0.64 & 10.31 &  0.98 &  1.00\\
    6 & 6886.6106 &  33.89  &    2.66  &    2.40 & 13.99 &  5.03 &  4.89\\
    7 & 7015.5107 &  27.38  &    0.45  &    0.42 &  8.81 &  0.73 &  0.77\\
    8 & 7076.3332 &  27.57  &    0.47  &    0.47 & 8.49  &  0.72 &  0.72\\
    9 & 7110.2534 &  22.68  &     1.21 &    1.03 & 9.84  &  2.05 &  1.99\\
    10& 7240.6384 &  27.18  &    0.54  &    0.50 & 9.32  &  0.86 &  0.90\\
    11& 7301.4733 &  28.35  &    0.49  &    0.47 & 8.55  &  0.77 &  0.74\\
    12& 7343.3616 &  28.01  &    0.51  &    0.50 & 9.46  &  0.92 &  0.91\\
    13& 7389.4760 &  29.85  &    0.52  &    0.49 & 10.26 &  0.76 &  0.74\\
    14& 7423.3999 &  26.78  &    0.43  &    0.45 & 8.08  &  0.84 &  0.85\\
    15& 7665.4718 &  28.73  &    0.67  &    0.64 & 8.08  &  0.84 &  0.85\\
    16& 7688.4124 &  26.79  &    0.38  &    0.37 & 9.84  &  0.71 &  0.72\\
    17& 7725.3120 &  23.91  &    0.59  &    0.52 & 3.98  &  0.84 &  0.82\\
    18& 7752.4980 &  26.84  &    0.48  &    0.58 & 7.77  &  0.95 &  0.837\\
    19& 7807.3459 &  26.98  &    0.47  &    0.46 & 8.38  &  0.87 &  0.86\\
    20& 7968.6508 &  25.85  &    0.43  &    0.47 & 6.70  &  0.84 &  0.83\\
    21& 8041.4353 &  26.06  &    0.90  &    0.81 & 7.70  &  1.40 &  1.38\\
    22& 8100.5350 &  28.87  &    0.35  &    0.35 & 10.10 &  0.62 &  0.60\\
    23& 8173.3314 &  27.70  &    0.49  &    0.43 & 9.42  &  0.77 &  0.74\\
    24& 8375.5344 &  25.53  &    0.35  &    0.35 & 9.29  &  0.68 &  0.68\\
    25& 8434.3638 &  24.22  &    0.37  &    0.36 & 8.21  &  0.65 &  0.63\\
    26& 8463.5479 &  22.23  &    0.48  &    0.44 & 6.87  &  0.74 &  0.83\\
  \end{tabular}
\end{table*}

The raw data were reduced by SALT staff, with the help of semi-automatic pipeline coming from the SALT PyRAF package\footnote{ http://pysalt.salt.ac.za}. Two exposures were combined, in order to increase the signal-to-noise ratio, and to remove the effect of the cosmic rays. The wavelength calibration was performed using the corresponding lamp exposure (argon lamp in most of the observations). One-dimensional spectra of the object and of the background were obtained using the IRAF package noao.twodspec. We used always the same width of the image for spectrum subtraction in order to reduce the unnecessary scatter.

SALT telescope has considerable vignetting effects, so the spectra required re-calibration with the help of a standard star, and for each observation we used the method described in detail by \citet{modzelewska2014} in their Section 2.1.

\subsection{Mg II line fitting}

We model the spectrum in detail in the 2700 \AA~ - 2900 \AA ~ wavelength frame, in the rest frame of the source. The decomposition of the spectrum has been done assuming the following components: power law (representing the emission from an acccretion disk), Fe II pseudo-continuum, and Mg II line itself. The power law is parameterized by its normalization and slope. For Fe II pseudo-continuum, we use one of the theoretical templates from \citet{bruhweiler2008}, d12-m20-20-5, which was favored for this source in the previous study \citep{modzelewska2014}. This template assumes the cloud number density $10^{12}$ cm$^{-3}$, the turbulent velocity of 20 km s$^{-1}$, and the flux of the hydrogen ionizing photons of $10^{-20.5}$ cm$^{-2}$ s$^{-1}$. The template was convolved with the Gaussian profile with dispersion of 900 km s$^{-1}$ representing the kinematic broadening of the Fe II lines. The assumed broadening and the choice of the Fe II template has been tested in \citet{modzelewska2014}.

\begin{figure}
  \centering
  \includegraphics[width=0.45\textwidth]{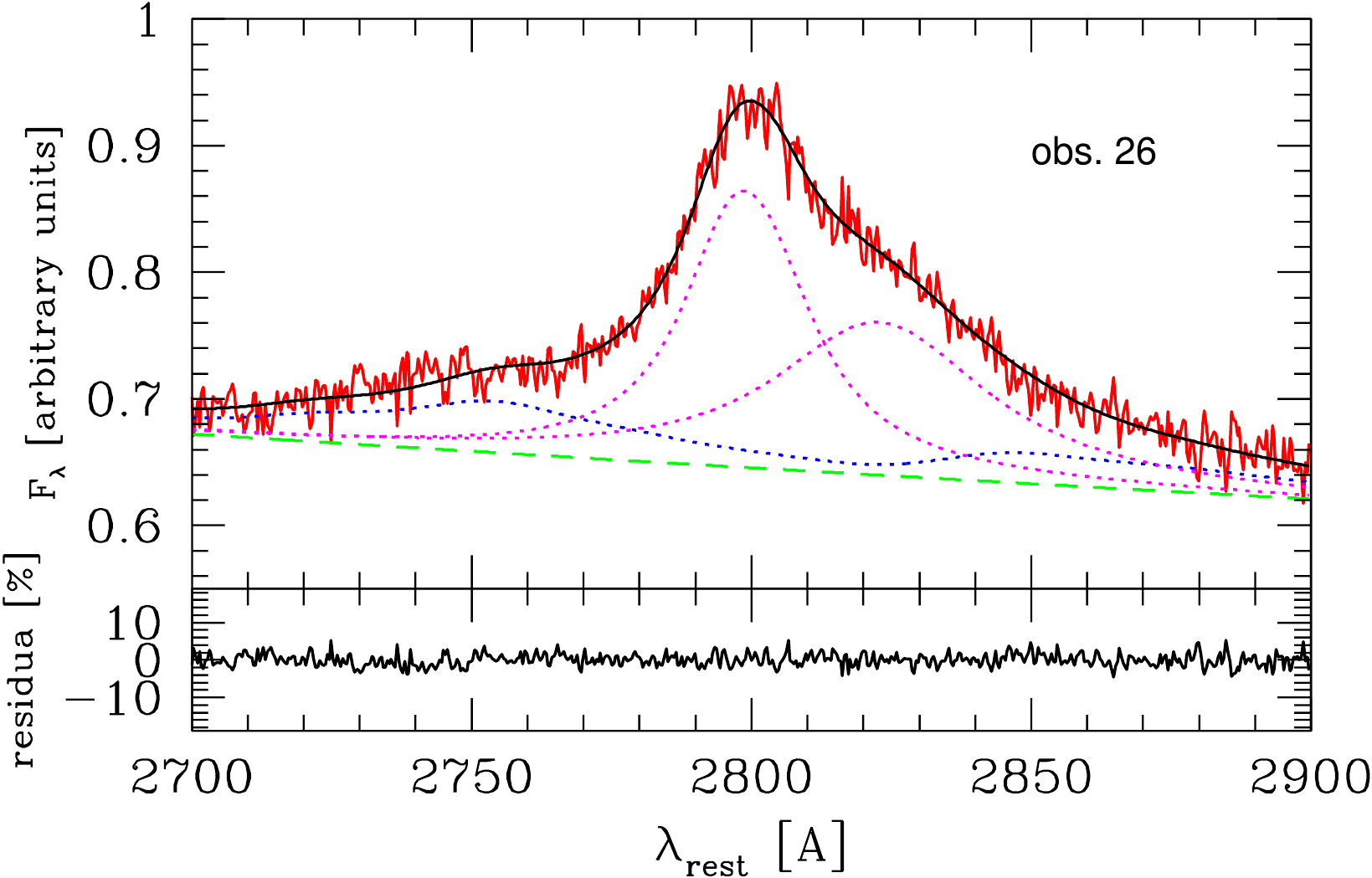}
  \caption{An exemplary SALT spectrum (last observation from 11 December 2018 - red continuous line) in arbitrary units, and decomposition into the underlying power law (dashed green line), Fe II pseudo-continuum (blue dotted line), and two broad kinematic components modelling the shape of the Mg II line (dotted magenta line). The total fit is shown as a continuous black line.}
  \label{fig:widmo_26}
\end{figure}

The Mg II line itself has been modeled using two kinematic components, as in \citet{modzelewska2014}, since a single component does not provide an adequate fit to the data. Each of the components is modelled assuming a Lorentzian shape, since this model minimizes the $\chi^2$ fit for a given number of free parameters. Gaussian components give higher values of  $\chi^2$ in most of the data sets. Each of the kinematic components is fitted as a doublet, with wavelenghths at 2796.35 \AA~ and 2803.53 \AA ~\citep{morton1991}. We used the doublet ratio 1.6 which provided the best fit in tested data sets. The best fit source redshift was derived to be 0.90052, and in our fits it corresponded to the blue kinematic component. The shift of the second (red) kinematic component was a free parameter of the model in each data set. The contribution of the Narrow Line Region is negligible in this source, and in \citet{modzelewska2014} we only derived an upper limit of 2 \%, so we did not include any component representing this emission. An exemplary SALT spectrum and its decomposition into the underlaying power law, Fe II pseudo-continuum and two kinematic components are shown in Fig.~\ref{fig:widmo_26}.

The total equivalent width (EW) of the line showed significant variations. The values of EW are listed in Table~\ref{tab:SALT_data}.  We also calculated the EW of the Fe II line in the spectral range 2700 - 2900 \AA, and we also give these values in Table~\ref{tab:SALT_data}.  The errors to EWs of Mg II and Fe II  were determined by construction of the error contours and determining the appropriate limit for one parameter of interest (other parameters during the fitting were allowed to vary), and they represent 90 \% confidence level.

\subsection{Photometry}
\label{sect:photo}

\begin{figure}
  \centering
  \includegraphics[width=0.45\textwidth]{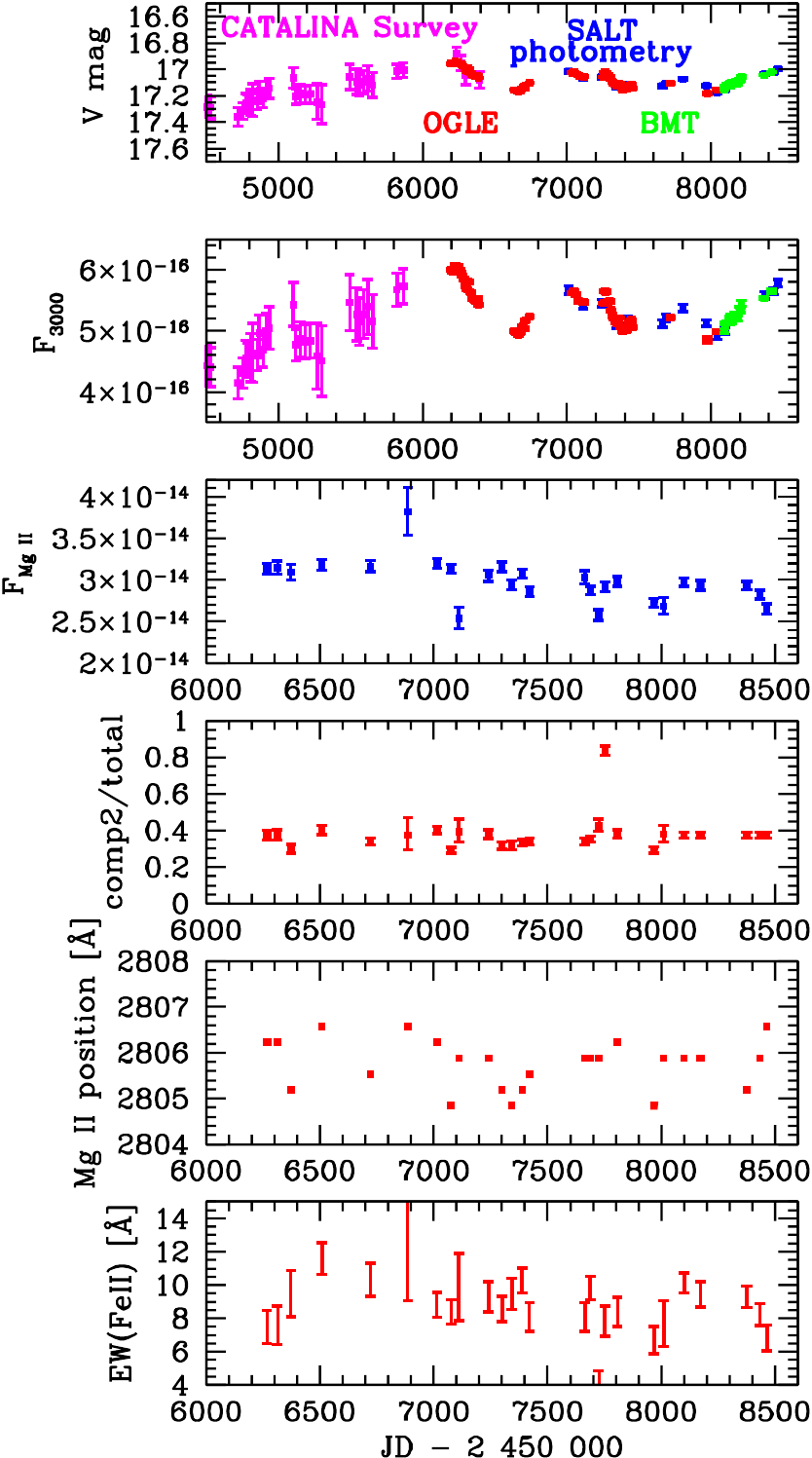}
  \caption{Photometric lightcurve of CTS C30.10 from three instruments (uppermost panel), flux at 3000 \AA~ in erg s$^{-1}$ cm$^{-2}$ \AA$^{-1}$, Mg II line flux in erg s$^{-1}$ cm$^{-2}$, the ratio of the second kinematic component of the line to the total line flux, the mean position of the line, and the EW of the Fe II pseudo-continuum measured between 2700 \AA~and 2900 \AA ~ (lowest panel). There is one clear outlier with large errorbars in the line flux plot (observation 6 when the weather conditions were not very good), and one outlier in the component ratio plot (observation 18; see Fig.~\ref{fig:widmo_18}). }
  \label{fig:curves}
\end{figure}

The spectroscopic observations were supplemented with the photometric data from other instruments, whenever possible. High quality photometry covering the first part of the observational campaign came from the OGLE-IV survey done with the 1.3 m Warsaw telescope at the Las Campanas Observatory, Chile. The exposures were done in the V-band, with the exposure time of 240 s. These data have typical error of 0.005 mag. The relative stability of this photometry for our source has been shown in \citet{modzelewska2014}.

At later epochs OGLE data was not available but the spectroscopic observations by SALT were usually performed with supplementing photometry with SALTICAM, in g band. The exposure time was 45 s, and two exposures were usually done. SALT/SALTICAM instrument is not suitable for high quality photometry.   Due to distortions of images taken by SALTICAM, a special, simple method for photometry has been developed. This method allows to partially reduce the occurrence of significant changes in the number of counts for the background and objects depending on the position in the image. The measurement error is larger in this case than from OGLE, typically of order of 0.012. Since the two photometric sets were done using slighly different filters, we allowed for an arbitrary shift in the SALT photometry, and the amount of shift was optimized in the time span where the two measurements overlapped.

CTS C30.10 has been also observed between 2nd December of 2017 and 1st April of 2018 employing the 40\,cm Bochum Monitoring Telescope (BMT) located at the Universit\''{a}tssternwarte Bochum, near Cerro Armazones in Chile (http://www.astro.ruhr-uni-bochum.de/astro/oca).
The observations were carried out with the Johnson broad band filter $V_j$ ($\lambda_{eff} = 550\,$nm and zero mag flux $f_0 = 3836.3$ Jy). Detailed information about the telescope, filters and data reduction has been published in \citet{2013AN....334.1115R}. The lightcurve for this period of time is built by choosing 20 stars, bright enough and in proximity of the source as calibration stars, the
aperture used for the photometry is $3.75 \arcsec$. The absolute photometric calibration is performed usually by comparison with the
PANSTARSS Catalog. Since CTS C30.10 is not available in the PANSTARSS catalog, then we used the conversion factor found for other objects that were observed with the same telescope, during the same nights. The photometry is given in Table~\ref{tab:photometry}.

Having the photometric coverage of the monitored period, we obtained the photometric lightcurve and, by linear interpolation, we also obtained the
Mg II lightcurve, ready for further analysis, including time delay measurements. The normalized dispersion of the continuum, in the whole period is 6.0\%, and the line flux variability, if we neglect observation 6, is at the level of 5.2 \%, lower (but not much) than the variability amplitude of the continuum.

In the period of 2 February - 14 March 2018 we also performed dense photometric observations of CTS C30.10 with the use of the 1.3 m SMART telescope at Cerro Tololo Inter-American Observatory. Photometry from these data was prepared using IRAF daophot software, and was typically of accuracy of 0.02, which is lower than what was achieved with other instruments, so we finally used these data  only to constrain the variability properties of the quasar in Sect.~\ref{sect:variability}.

\section{Spectral properties in the Mg II line region}

\begin{figure}
  \centering
  \includegraphics[width=0.45\textwidth]{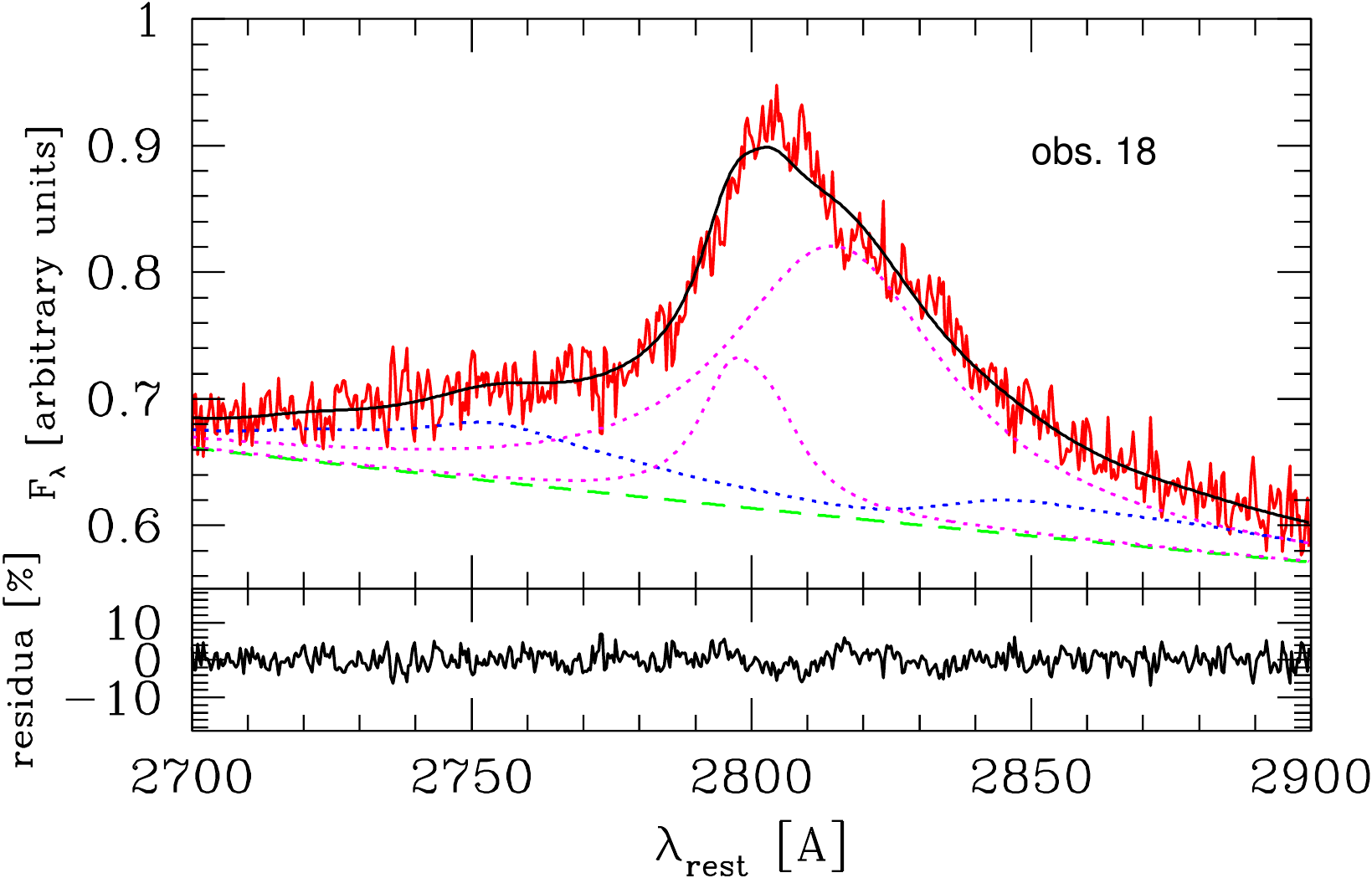}
  \caption{The SALT spectrum (observation 18, 29 Dec 2016) which shows atypical ratio of the two kinematic components of the Mg II line seen as a strong outlier in Figure~\ref{fig:curves}. See Figure~\ref{fig:widmo_26} for the description of spectra decomposition.}
  \label{fig:widmo_18}
\end{figure}

Since the Mg II line fitting in CTS C30.10 requires two broad kinematic components, we first analyze the variability in the spectral shape of the line reflected directly in the line fitting. In Fig.~\ref{fig:curves} we show the time dependence of the ratio of the second Lorentzian kinematic component to the total line.

One of the points, corresponding to Observation 18, is a strong outlier, with the second component strongly dominating. The value of the $\chi^2$ in this fit is particularly high, but we were not able to find another fit, with typical value of the component ratio. The overall spectrum in this observation (see Figure~\ref{fig:widmo_18}) is not very different from the representative spectrum (see Figure~\ref{fig:widmo_26}), but there seems to be a trace of absorption close to the peak of the line, which leads to such atypical spectral fit. Apart from observation 18, the mean FWHM of the first component is 2784 km s$^{-1}$, and the second component is slightly broader, with mean FWHM of 3884 km s$^{-1}$.

There is an overall weak linear rising trend in the ratio, but this might be related to the fact that we fixed the redshift of the first kinematic component, and if there is a systematic change in the Mg II line position as a whole line, it would show this behavior. Line shifts in quasars is a general phenomenon observed in high quality data \citep[see e.g.][]{sredzinska2017}. We thus test the line position in the data, by subtracting from the total flux the fitted continua (power law and Fe II), and integrating the line flux over the wavelength up to a wavelength value where half of the total line luminosity is contained. This method is less sensitive to the accuracy of the peak determination of the line, and seems more appropriate for an asymmetric line. The resulting line position is shown in Figure~\ref{fig:curves}, and no systematic trend is seen in the case of CTS C30.10.

\subsection{Mean and rms spectrum}

\begin{figure}
  \centering
  \includegraphics[width=0.45\textwidth]{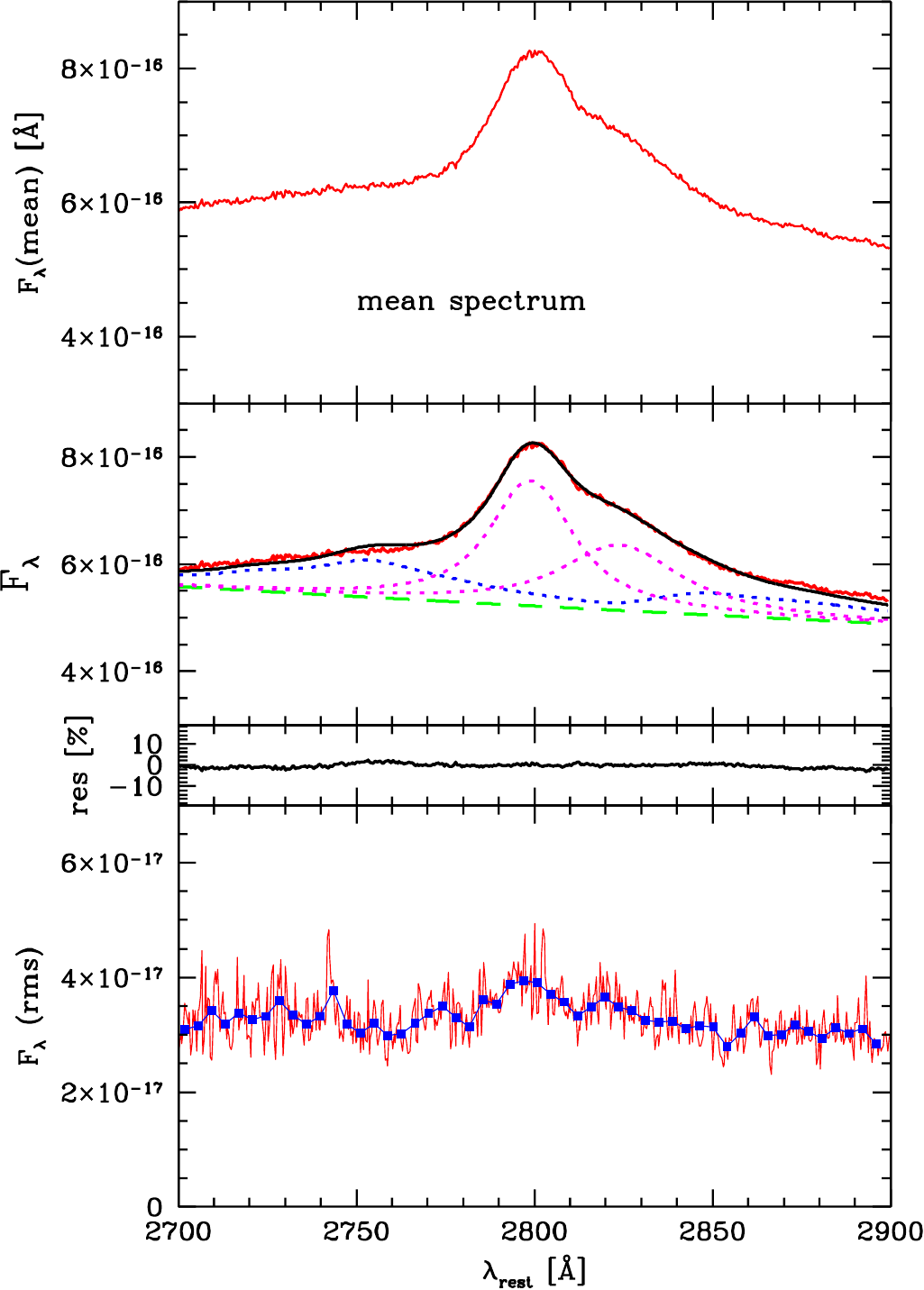}
  \caption{Mean (upper panel) spectrum, its decomposition (middle panel) and rms spectrum (lower panel) of CTS C30.10, calculated without observation 6.}
  \label{fig:aver}
\end{figure}

Since no obvious systematic drift is seen in the Mg II line, we calculate the mean and the rms variation of the Mg II line. For that purpose, we normalize each of the SALT spectra using the photometry provided in
Table~\ref{tab:photometry}. We had to remove observation 6 from the composite plot since this data is particularly noisy, flux varied up to a factor 2 between the consecutive wavelength bins, and that considerably affected the mean and rms. This observation was performed in the presence of some thin clouds. The remaining spectra are of much higher quality.

The mean spectrum traces the general shape of the Mg II line very nicely, with very low noise (see Figure~\ref{fig:aver}). The two-component character of the line shape is well seen. The mean spectrum is well fitted by the same model, there are some departures between the data and the model in the Fe II region, but other Fe II templates do not provide a better fit. Comparable (but slightly worse) fit was achieve by the Fe II template d11-m20-21-735 with lower number density ($10^{11}$ cm$^{-3}$) but higher ionizing flux. Using the mean spectrum decomposition, we measured the total FWHM of the Mg II line, treated as a single-component asymmetric line. The obtained value, 5009 km s$^{-1}$ locates CTS C30.10 among type B quasars in classification by \citet{sulentic2000}.

We removed observation 6 spectrum only in the mean and rms computations. Apart from that, this observation despite relatively high noise level gave meaningful value for the line flux (but with large error; see Fig.~\ref{fig:curves})  and was used in further considerations.

\section{Time delay measurements of Mg II line}
\label{sect:MgII_delay}

The number of spectra collected by us with the SALT telescope (26) just matches the minimum (25 spectra) necessary to obtain a viable estimate of the Mg II time delay \citep{czerny2013}. Therefore, we use five different methods to obtain reliable results. The first method is the classical Interpolated Cross Correlation Function (ICCF) method, most popular in AGN monitoring \citep[e.g.][]{gaskell1987,peterson2004}. \citet{edelson2018} argued that this method brings the most reliable results despite giving relatively large errors. Next we use the Discreet Correlation function. The third method is based on the Damped Random Walk (DRW) model of the quasar variability \citep{2009ApJ...698..895K}, and it was developed into the software JAVELIN \citep{2011ApJ...735...80Z,2013ApJ...765..106Z,2016ApJ...819..122Z}. We also use the method based on curve shifting and $\chi^2$ fitting, which was also used by several authors \citep[e.g.][]{kundic1997,czerny2013}. Finally, we use von Neumann estimator.

\subsection{ICCF}
\label{sect:ICCF}

\begin{figure}
  \centering
  \includegraphics[width=0.45\textwidth]{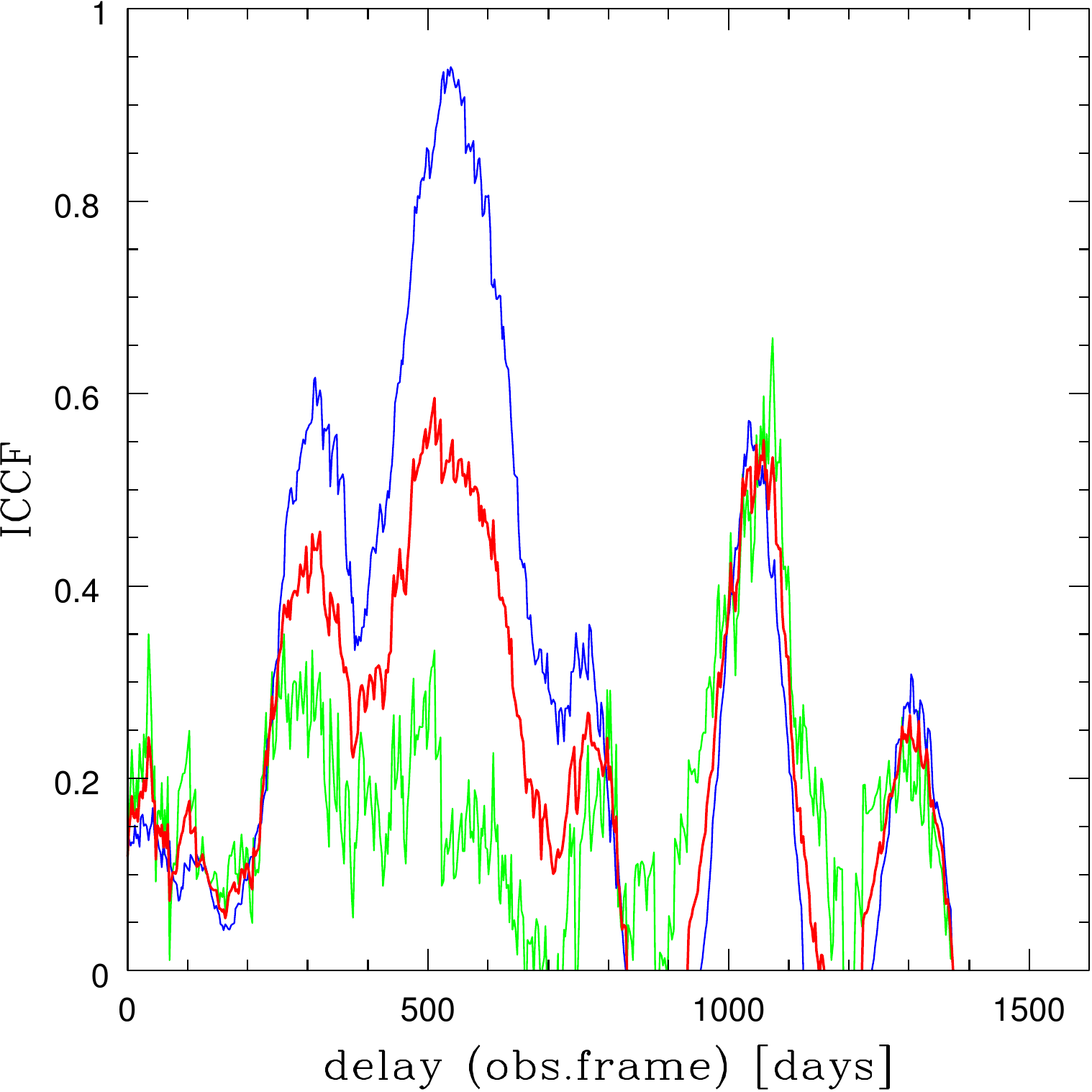}
  \caption{ICCF calculated by interpolating only photometry (green line), interpolating only spectroscopy (blue line), and the averaged symmetric result (red line).}
  \label{fig:iccf}
\end{figure}

Constructing ICCF is the traditionally used method of calculating the time delay in reverberation mapping of AGN \citep{peterson2004}. We mostly followed this general approach. However, we do not supplement the data curves with extrapolation when photometry or spectroscopy cannot be obtained through interpolation. \citet{peterson2004} argue that the extrapolation decreases the error but our lightcurves, particularly the spectroscopic one, are so short (only 26 measurements) that introducing not fully justified points may strongly bias the results.  Otherwise, we perform the computations in the standard way, subtracting the mean and normalizing the lightcurves by the variance, but at that stage we use the measurement errors, so our mean and variance are calculated as a weighted quantities. We think this is important since the errors in our data are not always similar to the typical error.

Next we perform ICCF. The best delay is determined by the fit of the centroid to the values above 0.8 of the peak values. As a basic output, we treat the solution obtained as a sum of two options: either photometric points are interpolated to match shifted spectroscopic points, or the spectroscopic points are interpolated to match the shifted photometric points.  However, we also use each of the two methods mentioned above separately, and they lead to strongly different results, shown in Figure~\ref{fig:iccf} and in the Table~\ref{tab:MgII_delay_nasze}. The asymmetric method, with the interpolation of spectroscopic points, leads to high ICCF cross-correlation values, $\sim 0.92$, close to the peak, implying the time delay of  542 days. On the other hand, interpolation of the photometric points shows much lower level of correlation, at most $\sim 0.51$, and the highest peak appears at 1073 days, and it strongly dominates the other peaks. Thus, high values of the ICCF in the previous case simply results from very few spectroscopic point being interpolated to dense photometric coverage which carry redundant information. Symmetrization points toward the lower value of the lag. There is also a clear second peak at 1070 days in the observed frame, but the ICCF peak is slightly lower. Thus, the result from ICCF is not unique, and this problem appears occasionally in reverberation measurements \citep[see e.g.][]{du2016}. However, taking into consideration that we use 133 photometric points and 26 spectroscopic points, it is not clear that averaging is more reliable that rather interpolating only the spectroscopic lightcurve. This would favor the longer value of the time delay, just above 1000 days. There is also a third peak, at 1300 days, and its height and position is similar in all three approaches.

The methods to obtain the errors of the lag determination were discussed by \citet{peterson1998}. This can be done by using the model-dependent or model-independent Monte Carlo method. We use here both methods.

We perform simulations assuming a shape and normalization of the power spectrum density (PDS) appropriate for CTS C30.10 (see Appendix). We use the general method of \citet{timmer1995} for generating the lightcurve in time domain when the PSD is known. The line lightcurve is produced assuming the shift by 294 days in the quasar rest frame for the case of the symmetric approach, and by smearing the line emission due to the extended BLR assuming a timescale equal 10\% of the time delay, using a Gaussian model of the BLR transfer function. Dense artificial lightcurves are then used to obtain continuum and line lightcurves for the actual observational cadence of our quasar. 1000 simulations were performed, and they were analysed using the same software as for the real lightcurves. The histograms created in this way were used as the probability distributions.  These errors are provided in Table~\ref{tab:MgII_delay_nasze} (as 1 $\sigma$ confidence levels). However, there is a long tail in the obtained distribution: in  81 of 1000 realizations the simulated time delay was longer than 500 days.

We also use the model-independent method based on bootstrap approach, as described in detail in \citet{peterson1998}. In this case the uncertainties are much larger, and the delay determination is highly uncertain, particularly the delay can be much longer than the best value. What is more, even the best fit value was affected by the change of the method. Quoting the bootstrap best value delay, we take it from the whole histogram, so both longer and shorter time delays are equally probable. This sensitivity of the results to the method is due to the fact that the distribution is not a single peak distribution, as clearly seen in Figure~\ref{fig:iccf} from ICCF distribution, and the same shows up in the histograms obtained through simulations when using the bootstrap method. In the previous, Timmer-Koenig based method, the assumed delay built into simulations were always recovered, and the histograms were never double-peaked.

Finally, we also tested the sensitivity of the results to the few spectroscopic points of likely lower quality. As a test, we removed observations No. 6, 9 and 17. All of them are clear outliers in Figure~\ref{fig:curves}, panel 3 (the first two cases) and the lowest panel (the last one, showing an apparent problem with Fe II fitting in these data). The results are also included in  Table~\ref{tab:MgII_delay_nasze}. The best time delay measurement favors even more strongly the longer time delay, about 1050 days in the observed frame. However, bootstrap modelling again bring results statistically concentrating around a shorter one.

\subsection{DCF}

\begin{figure}
  \centering
  \includegraphics[width=0.45\textwidth]{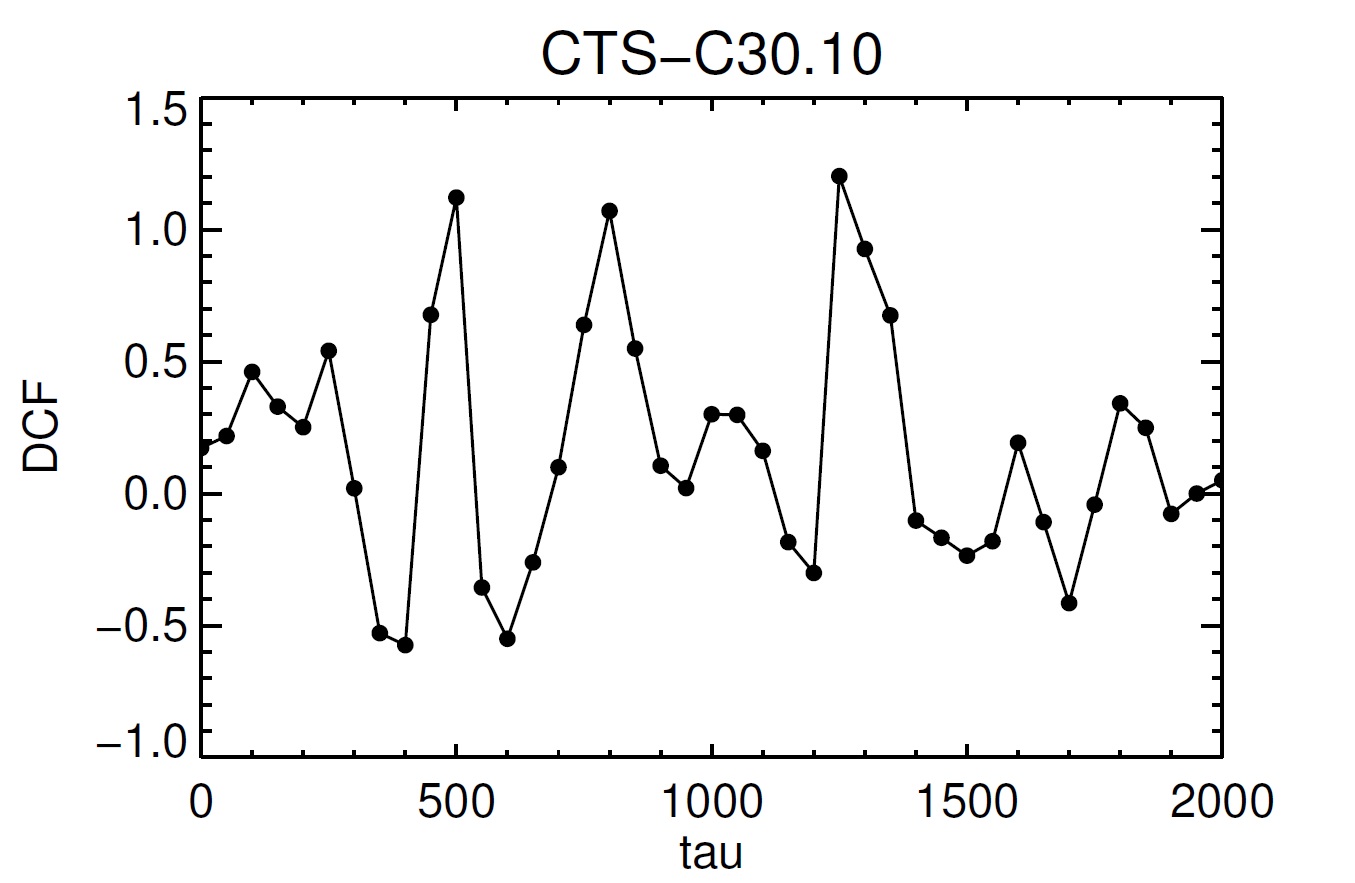}
  \caption{DCF calculated neglecting the observations 6, 9 and 17.}
  \label{fig:dcf}
\end{figure}

Since linear interpolation between points is not justified, Discreet Correlation Function (DCF) which does not require such interpolation has been proposed in the time delay measurement context by \citet{edelson1988}. We thus used this approach as well, removing the same three outliers as before, i.e. observations 6, 9 and 17. We also neglected the photometry before JD 245600 and after JD 2457200 to reduce the more noisy (first part) and the likely not needed part of the data. the resulting plot is shown in Figure~\ref{fig:dcf}. Again, we see a few peaks but the dominant peak is located at even longer timescales than ICCF best value of the time delay, at $\sim$ 1250 days.

\subsection{Javelin}
\label{sect:Javelin}

To measure the time-lag between the continuum emission and the line's response, we also use the
Javelin\footnote{\tt https://bitbucket.org/nye17/javelin} software (\citealt{2011ApJ...735...80Z,2013ApJ...765..106Z,2016ApJ...819..122Z}).
It models the continuum light curve as a stochastic process, here the damped random walk (DRW; \citealt{2009ApJ...698..895K,2010ApJ...708..927K,2010ApJ...721.1014M}), and assumes that the light curves of the emission line are the lagged, smoothed, and scaled versions of the continuum light curve. The Monte Carlo Markov Chain (MCMC) posterior probabilities produced by Javelin include such parameters as the two DRW model parameters (the variability time scale and amplitude), the time-lag, the smoothing width of the top-hat function, and the amplitude scale factor of the emission line to the continuum ($\rm{A_{line}/A_{photo}}$).

\begin{figure}
  \vspace{0.5cm}
  \centering
  \includegraphics[width=0.45\textwidth]{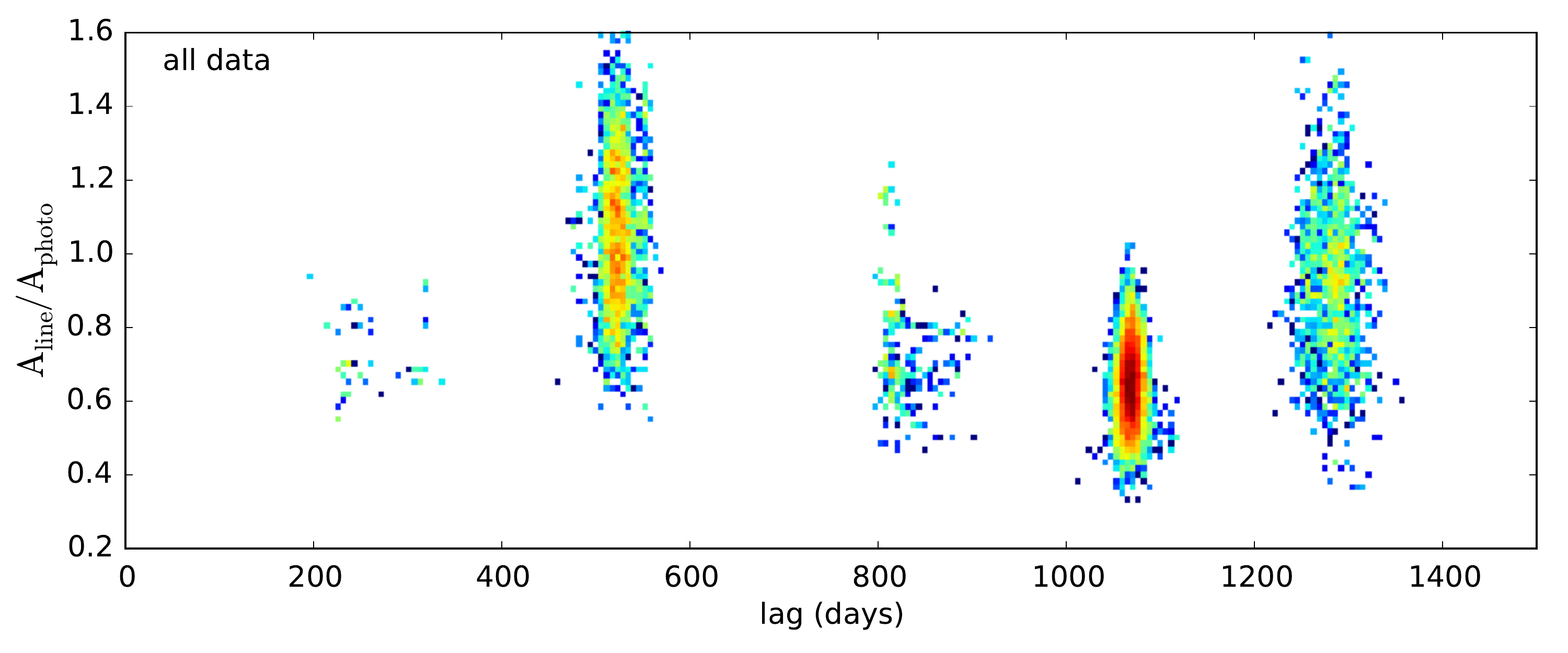}\\
  \includegraphics[width=0.45\textwidth]{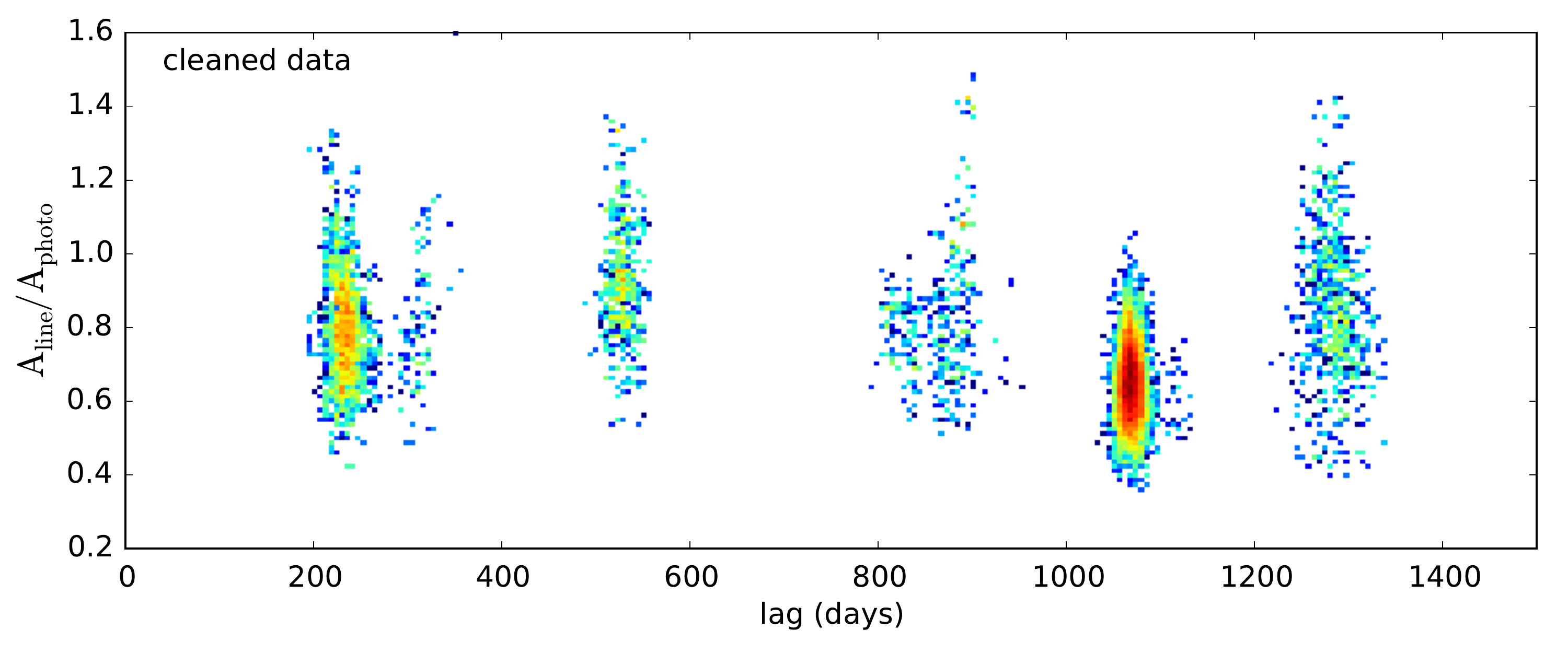}
  \caption{Posterior probability distributions obtained from the Javelin software for the full data (top) and cleaned line data (bottom). The strongest probability in both cases appears to be at the lag of 1068 days, that corresponds to 562 days in the rest-frame.}
  \label{fig:javelin}
\end{figure}

In Figure~\ref{fig:javelin}, we present the posterior probability distribution in the lag--$\rm{A_{line}/A_{photo}}$ plane. Several probability maxima are readily visible, of which the one at 1068 days is the strongest. An eye inspection of the line light curve uncovered two plausible outliers at ${\rm JD = 245 6886.5}$ and $245 7110.5$. We considered two cases here, where the line data were modeled in full (top panel in Figure~\ref{fig:javelin}) and with the two outliers removed (bottom panel of Figure~\ref{fig:javelin}). The best-measured observed-frame time-lags obtained from the strongest peak are $\tau=1067.85_{-3.63}^{+3.45}$ days for the full line light curve and $\tau=1067.64_{-3.42}^{+3.74}$ days for the cleaned light curve, where the error bars are asymmetric $1\sigma$ uncertainties. The rest-frame lags are then, respectively, $\tau=561.76_{-1.80}^{+1.97}$ days and $\tau=561.87_{-1.91}^{+1.82}$ days, fully consistent with each other.

In order to understand the probability structures in Figure~\ref{fig:javelin}, we performed simulations of 100 light curve pairs for (1) the input time-lag of 1000 days and (2) the input time-lag of 500 days. We used the DRW model with the typical input AGN variability parameters, the timescale of 1 year and the asymptotic variability of 0.25 mag (\citealt{2016ApJ...826..118K}). The cadence and photometric noise were set to be identical to the real data. Then we modeled the simulated data with Javelin in the same fashion as the real data, including the cadence and the errorbars.

We find that for the simulations with the lag of 1000 days, 10 cases (10\%) show incorrect measurements, other than the input 1000 days. In 25 (25\%)
cases we also find secondary probability peaks in the vicinity of 500 days (similar to what is observed in real data), 44 cases (44\%) probability peaks at 1200-1300 days, 27 cases (27\%) peaks at about 800 days, and 22 cases (22\%) peaks at about 200 days. Generally, the probability structures in the simulated data resemble quite well what we observe in the real data, and the additional peaks to the main one at 1000 days are expected based on simulations (even if there is a single lag present in the data).

Analysis of simulations with the 500 day lag, returns only 2 cases (2\%) with peaks in the vicinity of 1000 days, 41 cases (41\%) at about 1200 days.
General inspection of the output appears to be different than for the real data and simulations with the input lag of 1000 days.

\subsection{$\chi^2$ method}
\label{sect:chi2}

\begin{figure}
  \centering
  \includegraphics[width=0.45\textwidth]{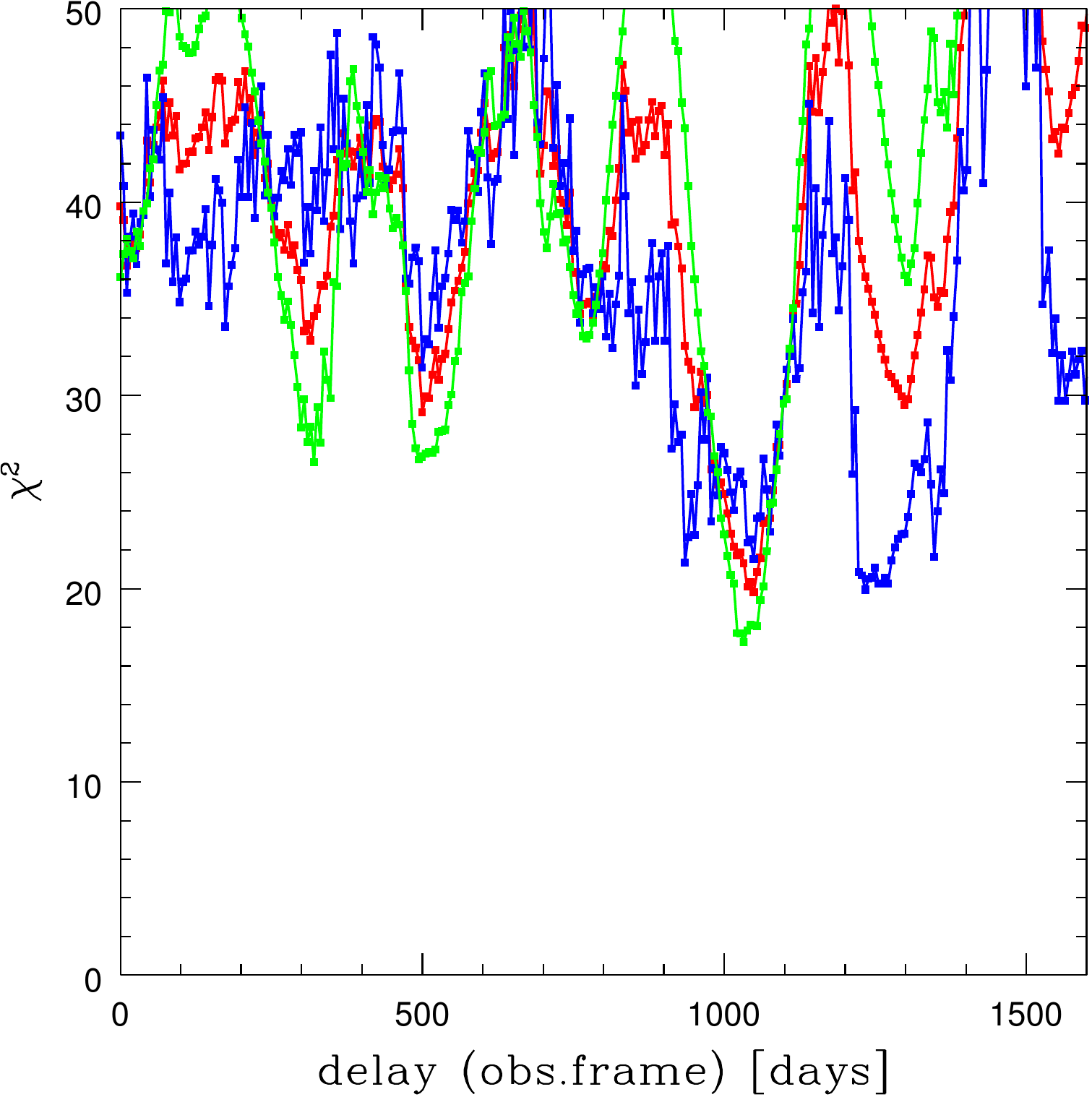}
  \caption{$\chi^2$ distribution, color-coded as in Figure~\ref{fig:iccf}, with minima implying the best solution for time delay.}
  \label{fig:moje_sym}
\end{figure}

This method is frequently used in quasar lensing studies, and \citet{czerny2013} found it works better than ICCF for the red noise AGN variability. Thus we also apply this method. The lightcurves are again prepared as for ICCF, by subtracting the mean and renormalizing by variance. Then the photometric or spectroscopic lightcurve is shifted with respect to the other one, and, as for ICCF, we can use the asymmetric approach (interpolating only photometry, or interpolating only spectroscopic lightcurve), or we can use both and average the results. We use only the linear interpolation, as in ICCF. We estimate the similarity of the shifted curves by simply calculating the $\chi^2$ value. Thus the minimum of the $\chi^2$ indicates the most likely time delay. The results are shown in Figure~\ref{fig:moje_sym}. The minimum is always at the location indicated by Javelin method (see Section~\ref{sect:Javelin}).

We assign the error again by performing Monte Carlo simulations as in Section~\ref{sect:ICCF}. The errorbars in the second column of Table~\ref{tab:MgII_delay_nasze} are from the PDS-based method, and in the third column we give the results from the bootstrap approach. The errorbars are much larger than in the case of Javelin method, comparable to ICCF. However, the distribution is more concentrated around the best solution, without an extended tail as in ICCF method. What is more important, the resulting most probable time delay is the same, independently from the choice of the data set for interpolation.

Also in this case we tested the sensitivity or the removal of the putlying spectroscopic points. We removed again three SALT observations No. 6, 9 and 17. The results for the best value of the time delay in this case did not change (see Table~\ref{tab:MgII_delay_nasze}), so the method seem quite stable. However, when boostrap modelling was used then the entire distribution of the possible time delays implies a lower value of the time delay, again with a very large error due to multiple maxima.

\subsection{von Neumann estimator}

\begin{figure}
  \centering
  \includegraphics[width=0.45\textwidth]{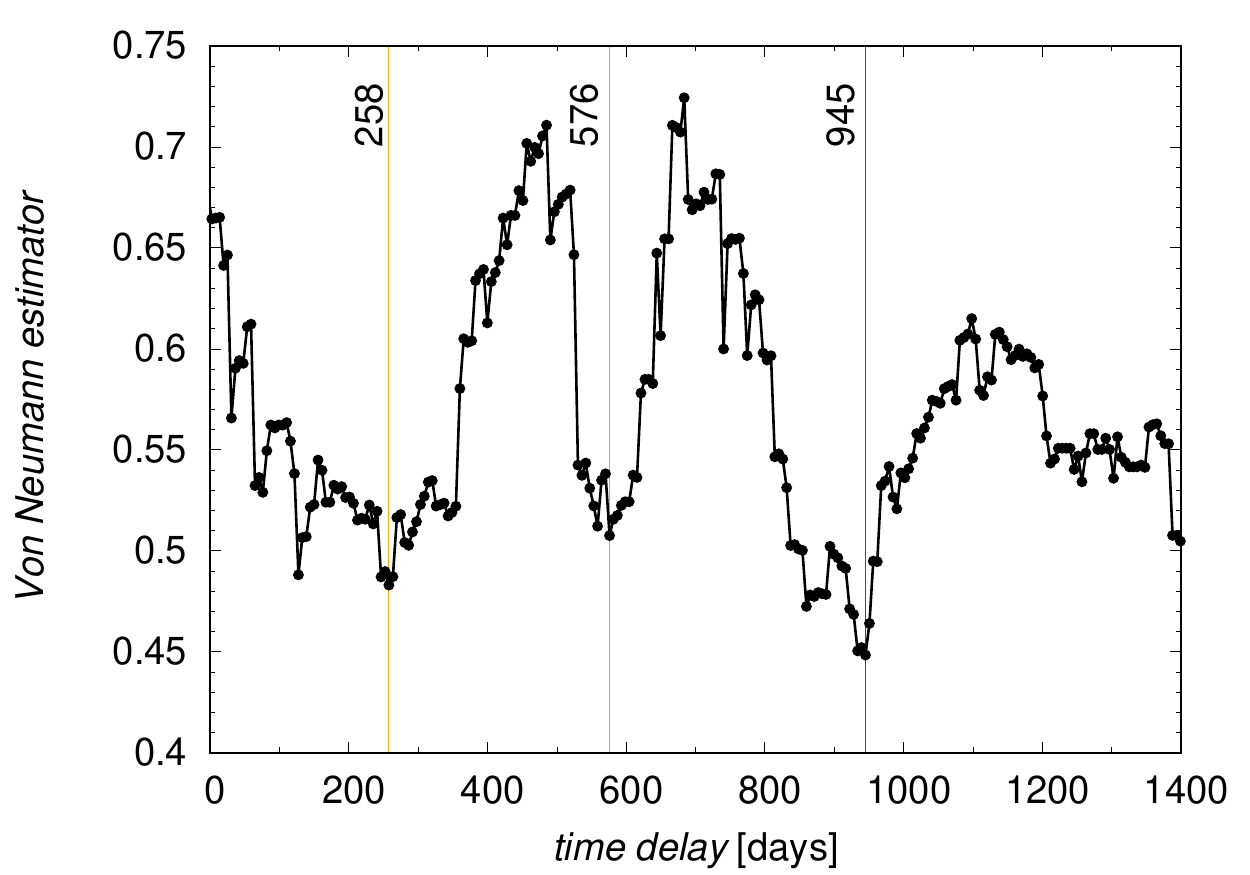}
  \caption{Dependence of the value of VN estimator on the time delay in days.}
  \label{fig_vn_estimator}
\end{figure}

Finally, we complement the previous ways of time-delay determination by the method of randomness measure or the complexity measure of the data. This method has an advantage that it does not require either the interpolation of data as for ICCF method or the binning in the correlation space and it is largely model-independent. The measures of the randomness or the complexity of the data are based on different kinds of estimators, among which the optimized von Neumann's scheme turns out to be most robust \citep{2017ApJ...844..146C}. The von Neumann (VN) estimator of the randomness (or regularity) of data is defined as the mean-square successive difference,

\begin{equation}
  E(\tau)\equiv \frac{1}{N-1} \sum_{i=1}^{N-1} [F(t_{\rm i})-F(t_{\rm i+1})]^2\,,
  \label{eq_von_neumann}
\end{equation}
where $F(t,\tau)=\{(t_{\rm i},f_{\rm i})\}_{i=1}^{N}=F_1 \cup F_2^{\tau}$ is the combined continuum and time-delayed line light-curve. When the searched time-delay $\tau$ is close to the actual time-delay between the light curves, $\tau \simeq \tau_0$, the VN estimator reaches the minimum. We applied the optimized scheme of the VN estimator based on the \texttt{python} script of \citet{2017ApJ...844..146C}\footnote{\url{http://www.pozonunez.de/astro_codes/python/vnrm.py}}. We show the dependence of the VN estimator on time-delay in Fig.~\ref{fig_vn_estimator}. In the search interval $[0,T_{\rm max}]=[0, 1300]$ days, there are three distinct minima, with the global one at $\tau_{0} \simeq 945$ days, followed by $\tau'=239$ days and $\tau''=573$ days. In the further analysis, we assume that the time-delay $\tau_0\simeq 945$ stands for the actual time delay, but this needs to be evaluated further statistically.

\begin{figure}
  \centering
  \includegraphics[width=0.45\textwidth]{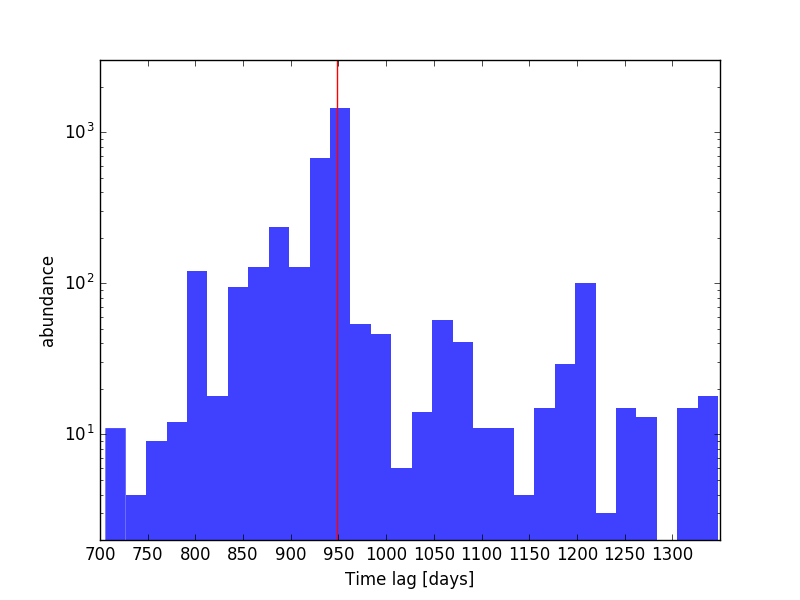}
  \caption{Distribution of the time-delay as determined by the VN estimator for $10\,000$ light curves generated by bootstrapping from the original observed pair of light curves. The red vertical line represents the mean value of $952$ days.}
  \label{fig_bootstrapping}
\end{figure}

To determine the uncertainty of $\tau_0$, we perform bootstrapping by generating subsamples of $10\,000$ continuum and line-emission light curves based on the actual observed light curves. This way we obtain a distribution of the time-delay around the global minimum, see Fig.~\ref{fig_bootstrapping}. The distribution is broad with the mean and the standard deviation of $\overline{\tau}_0=952\pm 90$ days, which is consistent within the uncertainty with the previously determined time-delays via the Javelin and $\chi^2$ methods, further strengthening the time-delay of $\sim 1000$ days. The advantage of the optimized VN method is that no stochastic model and assumption of quasar variability is needed.

\subsection{Best determination of the time delay}
\label{sect:best}

The quality of our data is not good enough to give always the same result, independently of the method. Three candidate time delay values appear: one around 1070 days, one at 550 days, and one at 1250 days in the observed frame. Javelin gives the smallest errors, and this method favors the intermediate delay, with or without two outliers removed.

The results from various methods are not easy to compare since error determination is very sensitive to the assumption about the quantity which error we measure. Javelin returns the error of the best solution position, while in other methods the bootstrap returns the errors based on the whole statistics, including side peaks as well. This basically explains the huge difference in errors between Javelin and $\chi^2$, ICCF and VN.

In the case of ICCF and $\chi2$ methods, the delay can be determined either using a symmetric approach to photometry and spectroscopy, or an asymmetric one.
Since the photometric coverage is much better, actually the asymmetric method requiring interpolation only for photometric data seems more justified.

Thus we calculated the average value of the best time delay by including only the full data ICCM and $\chi^2$ for photometric interpolation and the results from the other three methods. Such an average value is 1073 days in the observed frame, not far from the Javelin result itself. However, such an averaging is not statistically viable (although informative), and determination of the error of this quantity is even more problematic. Simulations with \citet{timmer1995} give errors of order of 60 days, but they do not include the information about the lightcurve properties, apart from the PDS. Bootstrap approach in the case of ICCF give values which seem too high, bootstrap approach to $chi^2$ indicates delays in the range from 939 days to 1281 days. Such an error is large but in our opinion this is the most realistic and conservative approach. So finally we adopt, as the best time delay, the value of 1074 days, and for the error we adopt the limits from the $chi^2$ bootstrap, i.e. $1073^{+208}_{-134}$. Javelin small error results are well within these limits.





\begin{table}
  \caption{Measured delays for the Mg II line in CTS C30.10 from various methods in observed frame}
  \label{tab:MgII_delay_nasze}
  \centering                          
  \begin{tabular}{l r r r}        
    \hline\hline
    Method & tau  & tau \\
    &    &  bootstrap \\
    \hline
    ICCF symmetric                    & $529^{+35}_{-30}$ & $537^{+1036}_{-274}$\\
    ICCF interp.phot.                 & $1073^{+39}_{-32}$ & $947^{+629}_{-401}$\\
    ICCF interp.spectr.               & $542^{+39}_{-60}$ & $ 288^{+1274}_{-34}$\\
    ICCF symmetric (-3 points) & 1041 & $412^{+224}_{-152}$ \\
    ICCF interp.phot. (-3 points) & 1064 & $926^{+27}_{-228}$ \\
    ICCF interp.spectr. (-3 points) &  569 & $318_{+225}^{-35}$\\
    \hline
    DCF  (-3 points)                & 1250 \\
    \hline
    Javelin all points                & $1067.9^{+3.5}_{-3.6}$ \\
    Javelin (- 2 points)      & $1067.6^{+3.7}_{-3.4}$ \\
    \hline
    $\chi^2$ symmetric    & $1049^{+40}_{-34}$ & $ 1191^{+101}_{-53}$\\
    $\chi^2$ interp.phot.   & $1032^{+60}_{-36}$ & $ 1148^{+133}_{-209}$\\
    $\chi^2$ interp.spectr. & $1233^{+48}_{-44}$ & $ 1182^{+42}_{-18}$\\
    $\chi^2$ symmetric (-3 points)     & 1038 &  $1185^{+75}_{-25}$\\
    $\chi^2$ interp.phot. (-3 points)  & 1022 & $1155^{+80}_{-202}$\\
    $\chi^2$ interp.spectr. (-3 points) & 1271 & $1203^{+102}_{-36}$ \\
    \hline
    VN                              & 945    & $948^{+90}_{-90}$ \\
    \hline
  \end{tabular}
\end{table}



\begin{table*}
  \caption{Measured delays for the Mg II line}
  \label{tab:MgII_delay}
  \centering                          
  \begin{tabular}{l r r r  r  r  r  r  r }        
    \hline\hline
    Object    &z  &   tau  &$err+$&  $err-$  &$\log L_{3000}$& err &Reference tau &Reference L \\
    &      & rest frame \\
    \hline
    141214.20+532546.7  &0.45810& 36.7& 10.4& 4.8& 44.63882  &0.00043& \citet{shen2016}     &\citet{shen2018}\\
    141018.04+532937.5  &0.46960& 32.3& 12.9& 5.3& 43.72880  &0.00506& \citet{shen2016}     &\citet{shen2018}\\
    141417.13+515722.6  &0.60370& 29.1& 3.6& 8.8& 43.68735  &0.00290& \citet{shen2016}     &\citet{shen2018}\\
    142049.28+521053.3  &0.75100& 34.0& 6.7& 12.0& 44.69091  &0.00090& \citet{shen2016}     &\citet{shen2018}\\
    141650.93+535157.0  &0.52660& 25.1& 2.0& 2.6& 43.77781  &0.00198& \citet{shen2016}     &\citet{shen2018}\\
    141644.17+532556.1  &0.42530& 17.2& 2.7& 2.7& 43.94795  &0.00105& \citet{shen2016}     &\citet{shen2018}\\
    CTS252    &1.89000& 190.0& 59.0& 114.0& 46.79373  &0.09142& \citet{lira2018}     &NED NUV GALEX\\
    NGC4151    &0.00332& 6.8& 1.7& 2.1& 42.83219  &0.18206& \citet{metzroth2006} &\citet{code1982}\\
    NGC4151    &0.00332& 5.3& 1.9& 1.8& 42.83219  &0.18206& \citet{metzroth2006}& \citet{code1982}\\
    CTS C30.10          &  0.90052  &   564.0    &   109    &  71     &    46.023$^*$       &    0.026       & this paper &  this paper \\
    \hline
  \end{tabular}
  ~~$^*$ obtained from the mean V mag 17.1, for the cosmology $H_0 = 70$ km s$^{-1}$ Mpc$^{-1}$, $\Omega_m = 0.3$, $\Omega_{\Lambda} = 0.7$ \citep{kozlowski2015}
\end{table*}\

\subsection{Radius-luminosity relation for Mg II line}

\begin{figure}
  \centering
  \includegraphics[width=0.45\textwidth]{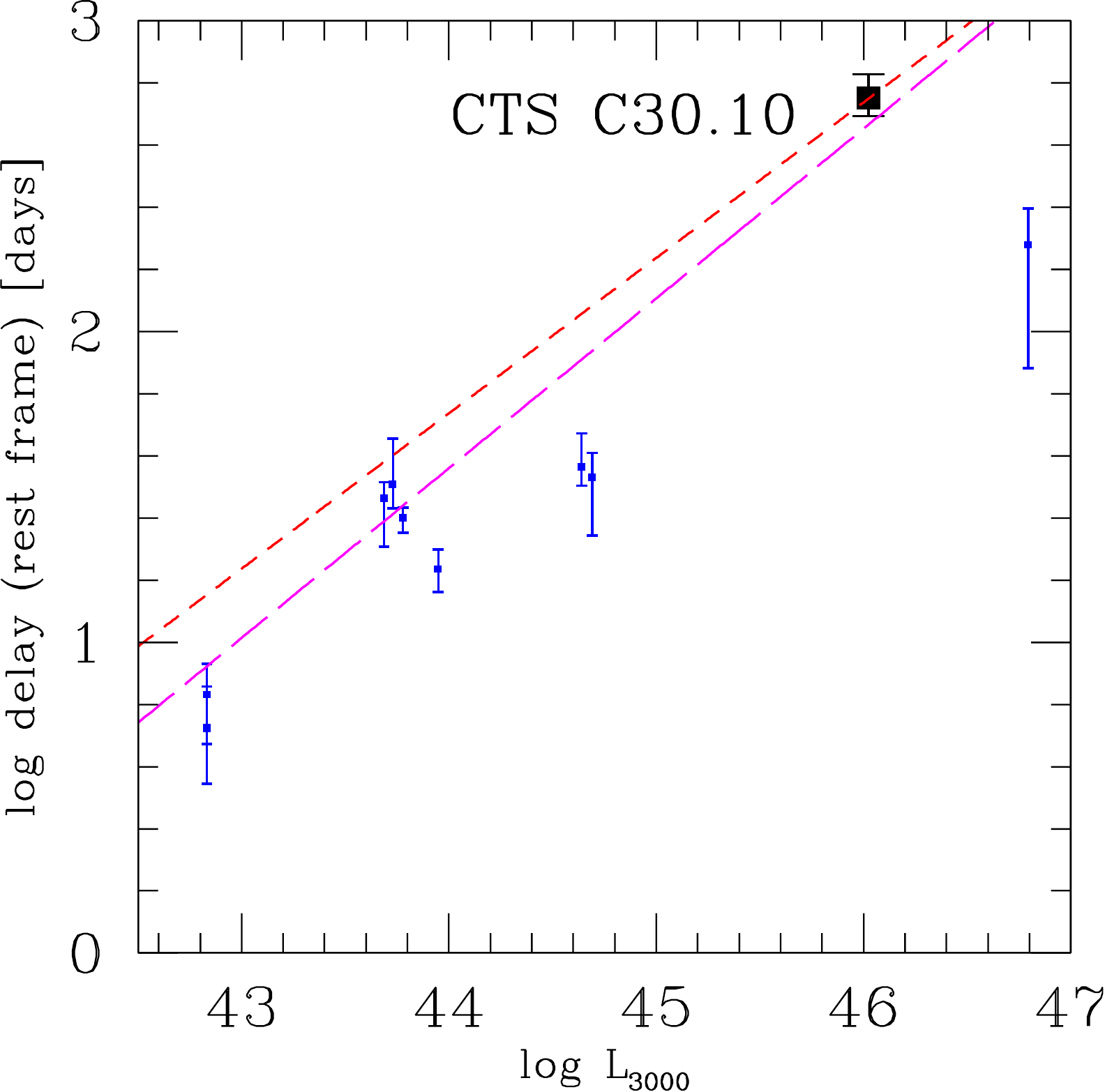}
  \caption{Radius-Luminosity relation for Mg II line (points and errors given in Table~\ref{tab:MgII_delay}). Dashed line is a line with the slope 0.5 passing through the measurement for CTS C30.10. Long dashed magenta line comes from the \citet{bentz2013} for H$\beta$ line (model Clean2) under the assumption that $L_{3000}$ = 1.84 $L_{5100}$.}
  \label{fig:R_L}
\end{figure}

We now combine the measurements available in the literature (see Section~\ref{sec:intro}) with our measurement of the time delay in CTS C30.10. The list of the measurements is given in Table~\ref{tab:MgII_delay}, and the plot using the final value set in Section~\ref{sect:best} is shown as Figure~\ref{fig:R_L}. Our source roughly follows the scaling with the slope 0.5 with the logarithm of the monochromatic luminosity at 3000 \AA, so a sequence similar to the sequence known from H$\beta$ line is apparently seen in Mg II line. CTS C30.10 is much brigher than AGN measured by \citet{shen2016}. On the other hand, it is somewhat less luminous and located at lower redshift than CTS252, but our time delay is longer than for CTS252. The black hole mass in CTS252 is $1.26 \times 10^9 M_{\odot}$, lower than the most probable mass in CTS C30.10 while the luminosity is higher by a factor of 5. The FWHM of the Mg II line is also much narrower in CTS252 (3800 km s$^{-1}$) than the total FWHM  of the line in CTS C30.10 ($5009 \pm 325$ km s$^{-1}$). This strongly suggests that the Eddington ratio in these two objects is different, much lower in CTS C30.10 than in CTS252. Thus the difference in the detected lag may be related to the difference in the Eddington ratios, in a similar way as it is for H$\beta$ line lags, with considerable lag shortening (for a given monochromatic luminosity) with the rise of the Eddington ratio (\citealt{du2016}; Martinez Aldama, in preparation).

As a reference, we plot in Figure~\ref{fig:R_L} the radius-luminosity relation of \citet{bentz2013}. Since now we use $L_{3000}$ on the x axis, we transformed the relation by recomputing $L_{3000}$ from $L_{5100}$ using the ratio of the bolometric corrections derived by \citet{richards2006}: $L_{3000}$ = (10.33/5.62) $L_{5100}$.



\section{Discussion}

Using our spectroscopic observations from our 6-year campaign performed with the SALT telescope and supplemented with photometry we were able to determine the time delay of the Mg II line in CTS C30.10.  Our result complements the delays measured for lower redshift lower luminosity sources.

We used five different methods to measure the time delay, and the Javelin method gave the best estimate, with the smallest error. Conservative error estimate, based on all methods give much larger error but the Javelin small error determination are well within the broader boundaries which we adopted in further study. However, it is interesting that apart from the strongly favored time delay just above 1000 days, there are strong traces of the secondary solutions at $\sim 500$ days, which even dominated in results from ICCF. Simulations based on PSD did not show such secondary peaks, but they were present in Javelin approach, and in all bootstrap results. The presence of the secondary peak does not necessarily mean that there are more than one timescale in the process but it might suggest that the transfer function is more complex than a simple Gaussian smearing adopted in \citet{timmer1995} simulations. The quality of our data is not good enough to test, if it is a signature of the broader range of timescales present in the system. For example the time delay for a two separate kinematic component may not be the same but we need more data points to test that.

Our measurement, combined with the previous results shows that the Mg II line time delay forms a similar pattern to H$\beta$ radius-luminosity (R-L) relation. The measured sources do not all follow the \citet{bentz2013}, but the same is seen in new measurements \citep{du2016,grier2017}, where the initially indicated R-L gave the maximum time delay for a given monochromatic luminosity bin. The Mg II line seems to be somewhat shifted (by some $\sim 20 - 50$ \% towards longer delays, but with just a few sources, and uncertain scaling between $L_{3000}$ and $L_{5100}$ we cannot give accurate value for this shift. \citet{wang2009} found that on average Mg II lines are narrower by some 15\% than H$\beta$ lines, which would imply an expected shift in the R-L by $\sim 25 $ \%, consistent with our finding. Similar shifts in black hole mass determination from H$\beta$ and Mg II lines were found by \citet{woo2018}.

The departure from the \citet{bentz2013} line might be caused by two effects: two short monitoring campaign or two high Eddington ratio of the source. The monitoring of \citet{shen2016} lasted 180 days in the observed frame, corresponding to about 125 days (z=0.45) for 141214.20, and about 100 days (z=0.75) for 142049.28. The campaign by \citet{lira2018} lasted about 10 years ($\sim $ 3650 days), hence the lag of 3757 days predicted from the R-L relation is larger than the monitoring campaign.
Thus in both cases the monitoring duration is likely too short
to properly measure the real delay, and we consider it very likely that
even the best cross correlation techniques are able to mimic too short lags
and too small errors implying high significance.

On the other hand, high Eddington ration sources show systematically shorter time delays than predicted by the \citet{bentz2013} relation \citep{du2016,du2018}. The source CTS252 \citep{lira2018} has high luminosity but relatively narrow lines implying low black hole mass implying close to Eddington, or even super-Eddington luminosity, while the most likely black hole mass based on the total width of the line in CTS C30.10 gives the Eddington ratio of order of a few per cent.
AGN with an Eddington ratio of a few per cent (like Seyfert 1 galaxies) tend to be much more variable in the optical band than high Eddington ratio sources, for example Narrow line Seyfert 1 galaxies \citep[e.g.][]{zhou2006}. Mg II line in our source is generally more variable than in most of the previously studied AGN. Also sources selected in the early reverberation monitoring were also highly variable. These sources now locate at longest time delays for a given luminosity, and the fact that our quasar also is characterized by a long time delay is not surprising.


\section{Conclusions}
\label{sec:conclusions}
Our main conclusions can be summarized as follows:
\begin{itemize}
\item We measured the rest-frame time-lag of $562\pm2$ days between the variable continuum and the Mg II emission line in a quasar CTS C30.10 ($z = 0.90052)$ using the Javelin method
\item We verified the reliability of this result by simulation means, where we explored the probability distributions in order to understand the multiple probability peaks, most of which turned out to be artificial
\item More conservative approach, based on a combination of five methods, leads to the rest frame time delay of $564^{+109}_{-71}$ days
\item{Mg II line time delay forms a radius-luminosity relation very similar to H$\beta$, with possible shift by less than 50\%. }

\end{itemize}

\section*{Acknowledgements}
The project was partially supported by National Science Centre, Poland, grant No. 2017/26/A/ST9/00756 (Maestro 9), and by the grant MNiSW grant DIR/WK/2018/12. S.K. and A.U. acknowledge the financial support of the Polish National Science Center through the grants: OPUS (2014/15/B/ST9/00093) and
MAESTRO (2014/14/A/ST9/00121). Polish participation in SALT is funded by grant No. MNiSW DIR/WK/2016/07, and the project is based on observations made with the SALT under programs 2012-2-POL-003 and 2013-1-POL-RSA-002,
2013-2-POL-RSA-001, 2014-1-POL-RSA-001, 2014-2-SCI-004, 2015-1-SCI-006,
2015-2-SCI-017, 2016-1-SCI-011, 2016-2-SCI-024, 2017-1-SCI-009,
2017-2-SCI-033, 2018-1-MLT-004 (PI: B. Czerny). VK acknowledges Czech Science Foundation No. 17-16287S. KH is grateful to Polish National Science Center for support under grant No. 2015/17/B/ST9/03422.

\facilities{SALT, SMART, OGLE, BMT, NED}

\software{Javelin \citep{2011ApJ...735...80Z,2013ApJ...765..106Z,2016ApJ...819..122Z}. Pyraf is the product of the Space Telescope Science Institute, which is operated by AURA for NASA. IRAF is distributed by the National Optical Astronomy Observatories, which are operated by the Association of Universities for Research in Astronomy, Inc., under cooperative agreement with the NSF.
}

\appendix
\section{Constraints to the character of the continuum variability and power density spectrum}
\label{sect:variability}

Javelin assumes DRW which requires the specific shape of the power spectral density (PSD). In order to assign errorbars in the ICCF and $\chi^2$ delay measurements with the use of model-dependent method we also need to know the PSD distribution in our source. We construct it with the help of the photometric data collected with SMART telescope (see Sect.~\ref{sect:photo} which cover the shortest timescales.

We calculate the normalized excess variance in the available timescales (2 months - SMART data, 6 years - combined OGLE and SALT photometry). We also divided the available lightcurves into two segments, calculated the excess variance separately for each segment, and averaged the result. That allowed us the access to additional time scales. Since OGLE photometry was particularly accurate, we also divided this lightcurve into 4 segments and averaged the results. The values are given in Table~\ref{tab:variance}.

\begin{table}
  \caption{Excess variance in the optical band for different timescales}
  \label{tab:variance}
  \centering
  \begin{tabular}{l r r}
    Instrument & Observation & Normalized \\
    & duration & excess variance \\
    \hline
    TOTAL & 4864.9& 3.78e-3\\  
    SALT+OGLE+SMART& 2233.3 & 3.56e-3 \\
    OGLE&1517.9 & 3.90e-3 \\
    SALT&1419.4 & 2.09e-3 \\
    SMART&70.9 & 3.85e-05\\
    \hline
    TOTAL/2 & 2432.45 & 2.86e-3 \\ 
    (SALT+OGLE+SMART)/2& 1116.6 & 3.15e-3\\
    OGLE/2&759.0 & 3.42e-3\\
    SALT/2&709.7 & 2.37e-3\\
    SMART/2&35.5 & 3.07e-6\\
    \hline
    OGLE/4 & 379.5 & 1.371e-3\\
    \hline
  \end{tabular}
\end{table}

Our data was not collected with the aim to recover full PSD accurately, but it interestingly compares with the PSD measurements in the optical band (\citealt{czerny1999} for NGC 5548; \citealt{czerny2003} for NGC 4151), \citealt{mushotzky2011} for four objects in Kepler data; \citealt{simm2016} for the sample of Pan-STARRS1 AGN, and \citealt{aranzana2018} for the large sample of Kepler AGN). The high frequency slopes in different objects span the range from -1.0 to -3.5, with the average value of -2.5 \citep{mushotzky2011,simm2016,aranzana2018}, and  -2.2 for detrended curves \citep{aranzana2018}. Similar high frequency slopes were found by \citet{2016ApJ...826..118K} using the Structure Function technique. The bends appeared at $\sim 100-300$ days in the rest frame,  with the average low frequency slope $\sim -1.1$ \citep{simm2016}. Some of their objects are close to the luminosity of CTS C30.10, but the frequency break in their study does not correlate with the black hole mass, but most of their observations last below 800 days in the quasar rest frame. Our data, including CATALINA extension, covers 2560 days in the quasar rest frame.

\begin{figure}
  \centering
  \includegraphics[width=0.45\textwidth]{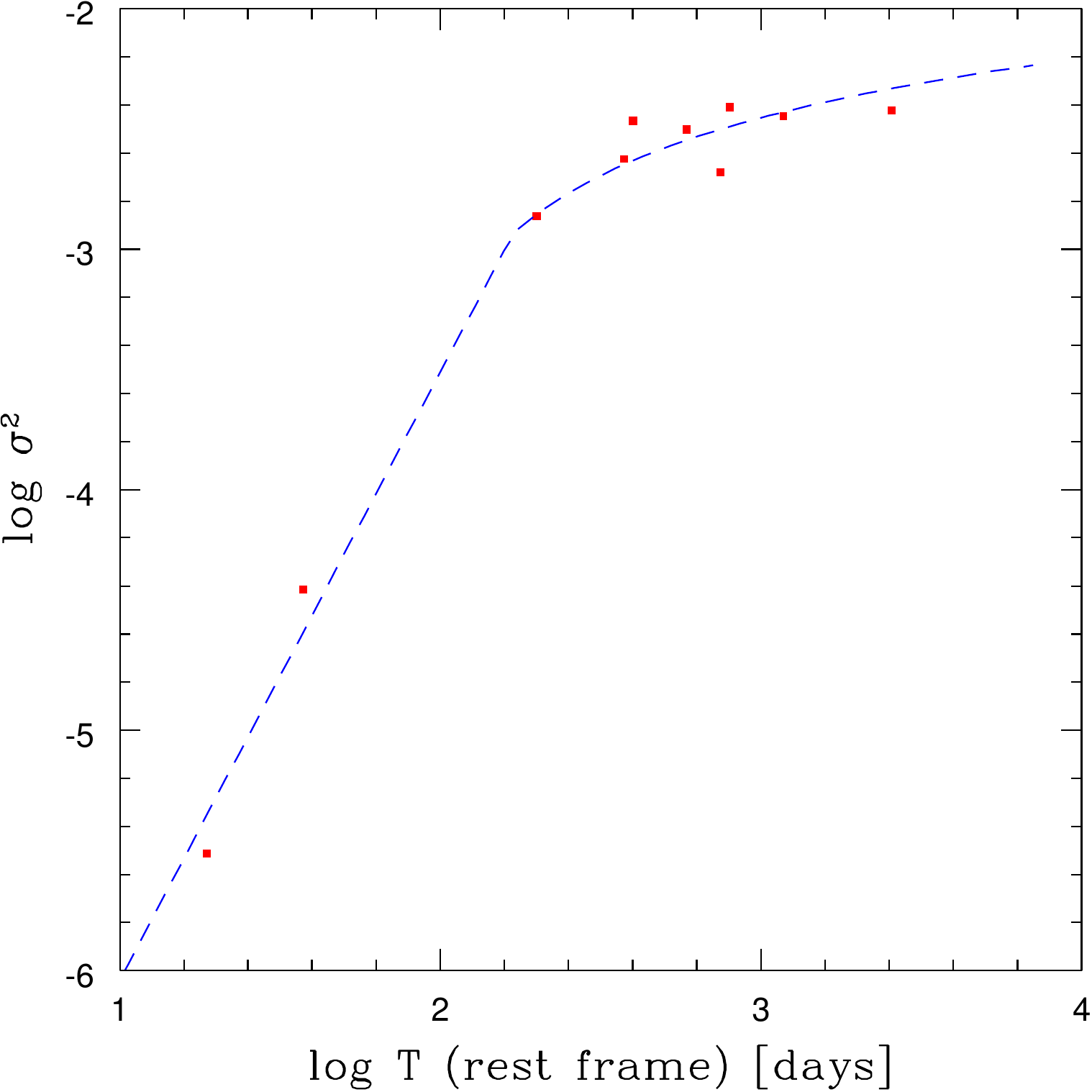}
  \caption{Excess variance of continuum (points) in the quasar rest frame as a function of duration of observations, together with the exemplary representation from the typical PSD shape of an AGN in a form of a broken power slope: low frequency slope 3.53, high frequency slope 1.064, frequency break corresponding to 168 days in the quasar rest frame (dashed line).}
  \label{fig:variance}
\end{figure}

Our excess variance from Table~\ref{tab:variance} is roughly consistent with this picture (see Figure~\ref{fig:variance}), as we clearly see the flattening of the excess variance rise. The flattening appears at $\sim 170$ days in the quasar frame. This timescale value is typical for AGN from \citet{aranzana2018} but much shorter than 2.5 yr timescale for NGC 5548 found by \citet{czerny1999}, particularly taking into account that the black hole mass in CTS C30.10 is larger (most likely $\sim 4.9 \times 10^9 M_{\odot}$, \citealt{modzelewska2014}) than the black hole mass in NGC 5548 ($6.54^{+0.26}_{-0.25} \times 10^7 M_{\odot}$, \citealt{bentz2007}).

We use this PDS to perform simulations which allow us to obtain the errorbars in Sections \ref{sect:ICCF} and \ref{sect:chi2}.

\bibliography{delays_CTS}
\bibliographystyle{aasjournal}

\startlongtable
\begin{deluxetable}{lrrl}
  \tablecaption{Photometric observations of CTS C30.10 performed with various instruments \label{tab:photometry}}
  \tablehead{
    \colhead{start of the observation}                &
    \colhead{V}     &
    \colhead{error}    &
    \colhead{Ref}                       \\
  }
  \startdata
  3597.75000&17.143&0.098&1\\
  3627.71875&17.203&0.067&1\\
  3667.70312&17.114&0.233&1\\
  3676.64844&17.231&0.100&1\\
  3692.68359&17.258&0.135&1\\
  3714.64062&17.196&0.079&1\\
  3733.61719&17.233&0.087&1\\
  3756.59766&17.255&0.078&1\\
  3819.46094&17.233&0.079&1\\
  3863.35938&17.309&0.078&1\\
  3980.78125&17.170&0.314&1\\
  4003.71875&17.258&0.087&1\\
  4013.71094&17.187&0.085&1\\
  4032.71875&17.244&0.128&1\\
  4049.66406&17.324&0.097&1\\
  4062.56641&17.203&0.090&1\\
  4079.61719&17.322&0.088&1\\
  4108.54688&17.294&0.091&1\\
  4119.54688&17.253&0.094&1\\
  4129.52344&17.267&0.088&1\\
  4174.48438&17.238&0.080&1\\
  4351.78125&17.250&0.089&1\\
  4394.68359&17.250&0.095&1\\
  4414.72656&17.238&0.103&1\\
  4442.62500&17.282&0.076&1\\
  4467.62109&17.260&0.076&1\\
  4503.46094&17.285&0.092&1\\
  4530.50000&17.290&0.081&1\\
  4721.73047&17.357&0.071&1\\
  4756.25000&17.314&0.065&1\\
  4775.72656&17.252&0.070&1\\
  4793.65625&17.262&0.070&1\\
  4808.56250&17.252&0.097&1\\
  4823.05859&17.234&0.122&1\\
  4855.57422&17.242&0.062&1\\
  4868.47266&17.160&0.067&1\\
  4894.48438&17.221&0.068&1\\
  4916.41016&17.154&0.066&1\\
  4939.39062&17.143&0.078&1\\
  5104.71875&17.063&0.075&1\\
  5121.71875&17.204&0.059&1\\
  5151.66406&17.183&0.067&1\\
  5188.60156&17.193&0.065&1\\
  5223.53906&17.188&0.066&1\\
  5269.43359&17.246&0.135&1\\
  5302.38281&17.266&0.148&1\\
  5495.71094&17.055&0.095&1\\
  5544.60156&17.095&0.095&1\\
  5561.61719&17.120&0.083&1\\
  5590.55469&17.096&0.085&1\\
  5618.52734&17.071&0.095&1\\
  5653.43750&17.120&0.096&1\\
  5820.76953&17.015&0.054&1\\
  5867.62500&17.005&0.056&1\\
  6199.79884&16.954&0.005&2\\
  6210.81712&16.960&0.004&2\\
  6226.67904&16.943&0.005&2\\
  6246.69777&16.945&0.004&2\\
  6257.74934&16.958&0.006&2\\
  6268.68328&16.962&0.004&2\\
  6277.68527&16.972&0.003&2\\
  6286.66883&16.984&0.005&2\\
  6297.61796&17.005&0.004&2\\
  6307.57569&17.014&0.004&2\\
  6317.64265&16.990&0.005&2\\
  6330.65842&17.022&0.004&2\\
  6351.54970&17.046&0.005&2\\
  6363.57515&17.050&0.004&2\\
  6379.48814&17.051&0.005&2\\
  6379.49575&17.045&0.005&2\\
  6387.51378&17.065&0.004&2\\
  6637.67203&17.154&0.004&2\\
  6651.62307&17.163&0.004&2\\
  6665.60641&17.167&0.004&2\\
  6678.60051&17.159&0.004&2\\
  6689.67477&17.136&0.004&2\\
  6700.63830&17.145&0.006&2\\
  6715.57763&17.117&0.004&2\\
  6740.49251&17.102&0.004&2\\
  7013.64000&17.016&0.012&3\\
  7036.65430&17.024&0.004&2\\
  7048.65618&17.021&0.004&2\\
  7060.60707&17.031&0.005&2\\
  7084.53744&17.052&0.005&2\\
  7110.91000&17.064&0.013&3\\
  7118.50958&17.055&0.005&2\\
  7240.35000&17.058&0.012&3\\
  7253.89503&17.058&0.004&2\\
  7261.88618&17.020&0.004&2\\
  7267.91792&17.021&0.005&2\\
  7273.85028&17.058&0.004&2\\
  7295.84577&17.052&0.005&2\\
  7306.78400&17.082&0.004&2\\
  7317.74327&17.101&0.005&2\\
  7327.77769&17.109&0.005&2\\
  7340.70964&17.126&0.004&2\\
  7342.33000&17.132&0.012&3\\
  7355.69767&17.119&0.005&2\\
  7363.66962&17.109&0.004&2\\
  7374.71221&17.138&0.004&2\\
  7385.56062&17.154&0.004&2\\
  7398.62071&17.145&0.004&2\\
  7415.58887&17.149&0.004&2\\
  7421.90000&17.113&0.012&3\\
  7426.56982&17.135&0.004&2\\
  7436.52896&17.123&0.005&2\\
  7447.53110&17.115&0.004&2\\
  7457.52592&17.140&0.004&2\\
  7664.25000&17.127&0.012&3\\
  7687.22000&17.107&0.012&3\\
  7717.70845&17.106&0.004&2\\
  7805.10000&17.075&0.012&3\\
  7965.86000&17.125&0.011&3\\
  7973.90556  &   17.181  &   0.008   &   4   \\
  7973.91644  &   17.188  &   0.010   &   4   \\
  8038.86464  &   17.155  &   0.006   &   4   \\
  8040.47000&17.171&0.012&3\\
  8091.20000&17.153&0.006&4\\
  8097.30000&17.125&0.006&4\\
  8099.30000&17.127&0.004&4\\
  8099.38000&17.153&0.012&3\\
  8128.20000&17.106&0.006&4\\
  8135.10000&17.102&0.006&4\\
  8142.10000&17.123&0.004&4\\
  8166.00000&17.097&0.006&4\\
  8174.00000&17.104&0.008&4\\
  8181.00000&17.110&0.008&4\\
  8197.00000&17.080&0.010&4\\
  8206.00000&17.082&0.008&4\\
  8211.00000&17.059&0.009&4\\
  8368.40000&17.043&0.004&4\\
  8374.40000&17.033&0.012&3\\
  8415.30000&17.020&0.005&4\\
  8433.05000&17.020&0.012&3\\
  8462.65000&16.995&0.011&3\\
  \hline
  \enddata
  \tablecomments{Ref.
    1: CATALINA survey,
    2: OGLE photometry,
    3: SALTICAM SALT photometry,
    4: BMT}
\end{deluxetable}

\end{document}